\documentclass[aps,twocolumn,showpacs,superscriptaddress]{revtex4-1}

\usepackage[pdftex]{graphicx}
\pdfoutput=1

\usepackage{amssymb,amsmath}
\usepackage{graphicx}
\usepackage{multirow}
\usepackage{color}
\usepackage[english]{babel}
\usepackage{hyperref}

\newcommand{\ket}[1]{ |{#1} \rangle }

\newcommand{\mtxelem}[3]{ \langle #1 \vert #2 \vert #3 \rangle }

\begin{document}

\def\simlt{\mathrel{\lower .3ex \rlap{$\sim$}\raise .5ex \hbox{$<$}}}
\def\simgt{\mathrel{\lower .3ex \rlap{$\sim$}\raise .5ex \hbox{$>$}}}

\title{\textbf{Probing low noise at the MOS interface with a spin-orbit qubit}}
\author{Ryan M. Jock}
\email[Corresponding author: ]{rmjock@sandia.gov}
\affiliation{Sandia National Laboratories, Albuquerque, NM 87185, USA}
\author{N. Tobias Jacobson}
\affiliation{Center for Computing Research, Sandia National Laboratories, Albuquerque, NM 87185, USA}
\author{Patrick Harvey-Collard}
\affiliation{Sandia National Laboratories, Albuquerque, NM 87185, USA}
\affiliation{D\'epartement de physique et Institut quantique, Universit\'e de Sherbrooke, 2500 boul. de l'Universit\'e, Sherbrooke, QC, J1K 2R1, Canada}
\author{Andrew M. Mounce}
\affiliation{Sandia National Laboratories, Albuquerque, NM 87185, USA}
\author{Vanita Srinivasa}
\affiliation{Center for Computing Research, Sandia National Laboratories, Albuquerque, NM 87185, USA}
\author{Dan R. Ward}
\affiliation{Sandia National Laboratories, Albuquerque, NM 87185, USA}
\author{John Anderson}
\affiliation{Sandia National Laboratories, Albuquerque, NM 87185, USA}
\author{Ron Manginell}
\affiliation{Sandia National Laboratories, Albuquerque, NM 87185, USA}
\author{Joel R. Wendt}
\affiliation{Sandia National Laboratories, Albuquerque, NM 87185, USA}
\author{Martin Rudolph}
\affiliation{Sandia National Laboratories, Albuquerque, NM 87185, USA}
\author{Tammy Pluym}
\affiliation{Sandia National Laboratories, Albuquerque, NM 87185, USA}
\author{John King Gamble}
\affiliation{Center for Computing Research, Sandia National Laboratories, Albuquerque, NM 87185, USA}
\author{Andrew D. Baczewski}
\affiliation{Center for Computing Research, Sandia National Laboratories, Albuquerque, NM 87185, USA}
\author{Wayne M. Witzel}
\affiliation{Center for Computing Research, Sandia National Laboratories, Albuquerque, NM 87185, USA}
\author{Malcolm S. Carroll}
\email[Corresponding author: ]{mscarro@sandia.gov}
\affiliation{Sandia National Laboratories, Albuquerque, NM 87185, USA}


\maketitle
\onecolumngrid

\section{Introduction}
\label{sec:introduction}

\textbf{The silicon metal-oxide-semiconductor (MOS) material system is technologically important for the implementation of electron spin-based quantum information technologies. Researchers predict the need for an integrated platform in order to implement useful computation \cite{Vandersypen2017}, and decades of advancements in silicon microelectronics fabrication lends itself to this challenge. However, fundamental concerns have been raised about the MOS interface (e.g. trap noise, variations in electron g-factor and practical implementation of multi-QDs).  Furthermore, two-axis control of silicon qubits has, to date, required the integration of non-ideal components (e.g. microwave strip-lines, micro-magnets, triple quantum dots, or introduction of donor atoms). In this paper, we introduce a spin-orbit (SO) driven singlet-triplet (ST) qubit in silicon, demonstrating all-electrical two-axis control that requires no additional integrated elements and exhibits charge noise properties equivalent to other more model, but less commercially mature, semiconductor systems \cite{Petersson2010, Shi2013,Dial2013,Wu2014}. We demonstrate the ability to tune an intrinsic spin-orbit interface effect, which is consistent with Rashba and Dresselhaus contributions that are remarkably strong for a low spin-orbit material such as silicon. The qubit maintains the advantages of using isotopically enriched silicon for producing a quiet magnetic environment, measuring spin dephasing times of 1.6 $\mathrm{\mu s}$ using 99.95\% $^{28}$Si epitaxy for the qubit, comparable to results from other isotopically enhanced silicon ST qubit systems \cite{HarveyCollard2015,Eng2015, Rudolph2017}. This work, therefore, demonstrates that the interface inherently provides properties for two-axis control, and the technologically important MOS interface does not add additional detrimental qubit noise.}

Creating qubits using the silicon metal-oxide-semiconductor (MOS) material system is compelling because of its quiet nuclear environment and the future promise of building upon CMOS capability. Recently, several critical demonstrations for qubit viability in MOS quantum dots (QDs) have been made including: (1) large tunable valley-splitting reproduced in multiple process flows \cite{Gamble2016}, (2) long spin coherence times \cite{Veldhorst2014}, and (3) fast coherent exchange coupling of spins in a multi-quantum dot layout \cite{Veldhorst2015a}. Yet, the intrinsically imperfect Si/SiO$_2$ interface produces persistent central concerns about charge traps and uncertainty in the degree to which the electron $g$-factor may be controlled \cite{Veldhorst2014}. In particular, charge traps and two-level fluctuators near the interface are believed to be potential sources of noise in MOS devices\cite{Ralls1984,Zimmerman2013}.  The prevailing material choice for Si qubits is heteroepitaxial Si/SiGe, which shifts the imperfect crystal-dielectric interface further away. However, there are doubts about reproducible valley splitting \cite{Borselli2011} in this system, and the qubit structures diverge from conventional CMOS design. No direct measurement of charge noise at the MOS interface has been made, though indirect measures on spin qubits \cite{Veldhorst2015a,HarveyCollard2015,Rudolph2017} suggest the effect will not impede coherent control. Variability in $g$-factor has also been observed, introducing complications in qubit device architecture \cite{Jones2016}.  Recent work has attributed variability in electron $g$-factor at silicon interfaces to spin-orbit coupling and interface disorder, including step edges \cite{Veldhorst2014,Veldhorst2015b,Ferdous2017,Ferdous2017b}. Spin-orbit coupling effects have been studied extensively in III-V quantum dot systems \cite{Pfund2007,Fasth2007, Nadj2010, Nadj2010b,Stepanenko2012,Rancic2014,Scarlino2014,Nichol2015}.  However, these effects have not, to date, been fully understood in the MOS material system. The results in this paper address both of these fundamental doubts about the MOS interface by showing that: (1) the spin-orbit interaction can be either turned off or tuned as a useful quantity, and (2) the charge noise at the MOS interface is comparable to other qubit systems.  

\begin{figure}[!h]
	\centering
	\includegraphics[width=0.7\textwidth]{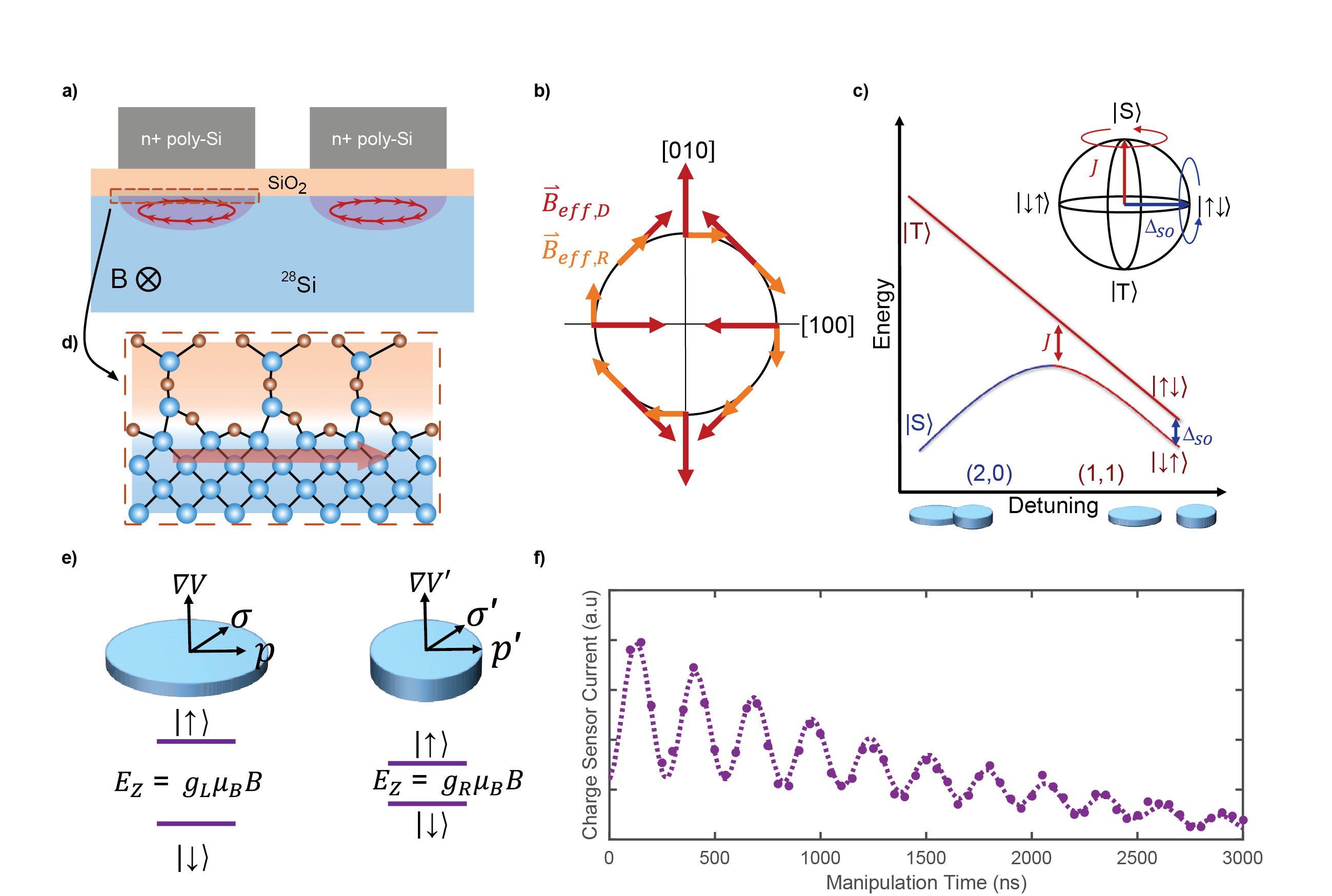}
	\caption{\textbf{MOS spin-orbit driven singlet-triplet qubit}. a) Cartoon representation of the interface spin-orbit interaction.  For an electron confined to a QD, an in-plane magnetic field will cause a finite momentum at the interface which, in the presence of broken inversion symmetry, leads to a spin-orbit interaction.  b) Schematic example of the effective spin-orbit field due to the Dresselhaus (red) and Rashba (orange) interactions for in-plane electron momentum. c) Schematic energy diagram of the DQD near the (2,0) $\rightarrow$ (1,1) charge transition, showing the energy of the singlet and triplet states as a function of QD-QD detuning, $\epsilon$. Near the interdot transition ($\epsilon$ = 0), the exchange energy, $J$, dominates the electronic interaction and drives rotations about the $Z$-axis (red arrow in inset). Deep into the (1,1) charge sector ($\epsilon$ $>$ 0), $J$ is small and the electronic states rotate about the $X$-axis due to a difference in Zeeman Energy between each QD (blue arrow in inset). d) Details of the interface at the inter-atomic bond level govern the spin-orbit interaction.  e) The local electrostatic environment of each QD leads to different momenta and electric fields at the interface and, thus, distinct spin orbit interactions and Zeeman energy splitting. f) Charge sensor current as a function of time spent deep in the (1,1) charge sector, where higher current indicates a higher probability of measuring a singlet state. The oscillations indicate clear $X$-rotations due to a difference in spin-orbit interaction in each QD.}
	\label{fig:Fig1}
\end{figure}

In bulk Si, the SO interaction leads to only weakly perturbed $g$-factors that are close to $g = 2.0$.  However, the inversion asymmetry of the crystal at an interface leads to a SO interaction\cite{Rossler2002,Golub2004,Nestoklon2006,Prada2011}, as shown in Fig. \ref{fig:Fig1}. When a magnetic field is applied with a component parallel to the interface, electron cyclotron motion establishes a non-zero net momentum component along the interface, Fig. \ref{fig:Fig1}(a). The coupling of the electron momentum perpendicular to the effective electric field at the interface produces the SO interaction. The asymmetry of the electric potential at the interface leads to a Rashba SO contribution due to structural inversion asymmetry (SIA). A second interaction, the Dresselhaus contribution, is attributed to microscopic interface inversion asymmetry (IIA)\cite{Nestoklon2008}.  The Rashba and Dresselhaus SO couplings for an electron confined to an interface have the form $H_{R} \propto \alpha_{R} \left( P_{y} \sigma_{x} - P_{x} \sigma_{y} \right)$ and $H_{D} \propto \beta_{D} \left( P_{x} \sigma_{x} - P_{y} \sigma_{y} \right)$, respectively, where $\alpha_{R}$ and $\beta_{D}$ are the relative coupling strengths. The operators $\sigma_{x}$, $\sigma_{y}$ are Pauli spin matrices, while $P_{x}$, $P_{y}$ are components of the kinetic momentum $\mathbf{P} = -i \hbar \nabla + e \mathbf{A}(\mathbf{r})$ along the $[100]$, $[010]$ direction, with $e>0$ the elementary unit of charge and $\mathbf{A(r)}$ the vector potential. This work quantitatively and unambiguously characterizes the SO effect at the MOS interface over its full angular dependence and provides a theoretical framework that removes the gauge-dependent ambiguity of previous models of interface Rashba-Dresselhaus coupling.  We note that this interface effect is not theoretically unique to Si interfaces \cite{Rossler2002,Golub2004,Alekseev2017}.  Because of its strength and angular dependence that is similar to bulk SO effects, it is possible that the contribution of the interface effect, particularly on the Dresselhaus coupling, is under-appreciated in other systems that leverage strong SO coupling.  Improved understanding of this effect has the potential to influence areas such as spintronics and the pursuit of forming new topological states of matter \cite{Manchon2015,Soumyanarayanan2015}.

\section{Results}
\label{sec:results}

The qubit in this work is formed within a MOS double quantum dot (DQD).  Two electrons are electrostatically confined within a double well potential, where the dominant interaction between the electrons can be electrically tuned between two regimes for two-axis control, Fig. \ref{fig:Fig1}(c).  When the electronic wave functions of the QDs overlap significantly, the exchange energy, $J$, dominates.  When the two electrons are well separated, $J$ is small and distinct Zeeman Hamiltonians result from the differences in their interface SO coupling.  The difference in SO coupling leads to a variation in effective electron $g$-factors, Fig. \ref{fig:Fig1}(e), and amounts to an effective magnetic field gradient between the QDs that can be tuned with control of the applied electric and magnetic fields.  Thus, we achieve all-electrical two-axis control using native features of the MOS DQD system, avoiding the substantial fabrication complications of other Si qubit schemes.

\begin{figure}[!h]
	\centering
	\includegraphics[width=0.9\textwidth]{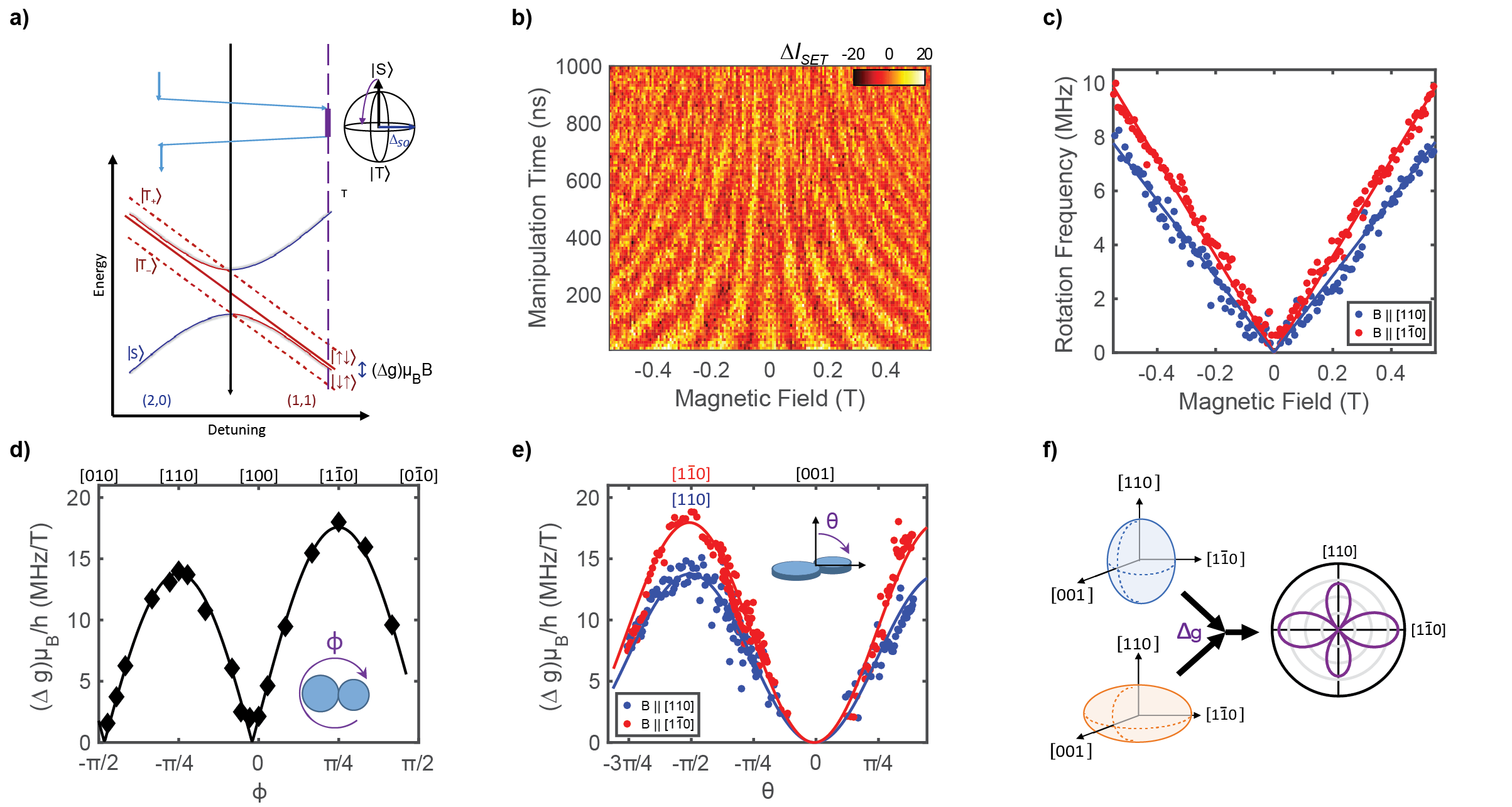}
	\caption{\textbf{MOS interface spin-orbit interaction}. a) Energy diagram and gate pulse schematic for controlling spin-orbit rotations. We initialize the qubit into the S(2,0) ground state and transfer the system to the (1,1) charge sector with a fast adiabatic pulse, such that it remains a singlet. The difference in Zeeman splitting between the QDs drives $X$-rotations between the $S(1,1)$ and $T_{0}(1,1)$ states.  A fast adiabatic return pulse projects the states onto the $S(2,0)$ and $T_{0}(1,1)$  basis for measurement. b) Change in charge sensor current as a function of $X$-rotation manipulation time as the magnetic field is varied along the $[1\bar{1}0]$ crystallographic direction. c) The extracted rotation frequency as a function of magnetic field strength along the $[110]$ and $[1\bar{1}0]$ crystallographic directions. d,e) Magnetic field angular dependence of the SO-driven difference in $g$-factor between the dots for the in-plane, $\theta$, and out-of-plane, $\phi$, directions, respectively. Fits to the form $(\Delta g) \mu_{B} B / h =\vert \mathbf{B} \vert \vert \Delta \alpha - \Delta \beta \sin(2 \phi) \vert \sin^{2}(\theta)$ are also plotted for $\theta = \pi/2$ (black), $\phi = 3 \pi/4$ (blue) and $\phi = \pi/4$ (red). f) A cartoon representation of the angular dependence of the two QDs (left). The difference between the QD $g$-factors give an in-plane dependence represented by the cloverleaf plot on the right.}
	\label{fig:Fig2}
\end{figure}

We define the computational basis as the eigenstates of the two-spin system in the limit of a large singlet-triplet exchange energy, $J$.  Specifically, these are the two states, $S$ and $T_{0}$, of the $m=0$ subspace, which form a decoherence-free subspace relative to fluctuations in a uniform B-field \cite{Lidar1998}.  An applied magnetic field splits the $m = \pm 1$ spin triplet states ($T_{\pm}(1,1)$) and $m = 0$ states by the Zeeman energy $E_{z} = g \mu_{B} B$ to isolate the $m=0$ subspace.  A qubit state can then be initialized in a singlet ground state when the two QDs are electrically detuned out of resonance such that it is preferable to have a $(\mathrm{N_{QD_{1}}}, \mathrm{N_{QD_{2}}}) = (2,0)$ charge state, Fig \ref{fig:Fig1}(c).  Rapid adiabatic passage to the $(1,1)$ charge state produces a superposition of the stationary eigenstates in the gradient field. A difference in the Larmor spin precession frequency of the two QDs induces $X$-rotations between the $S(1,1)$ and $T_{0}(1,1)$ states, Fig. \ref{fig:Fig1}(f) and \ref{fig:Fig2}(a). For each QD the angular precession frequency is given by $\omega = g\mu_{B}B/\hbar$, where $g$ is the electron $g$-factor, $\mu_{B}$ is the Bohr magneton, $\hbar$ is Planck's constant, and $B$ is the applied magnetic field. The two-electron spin qubit will oscillate between the $S$ and $T_{0}$ states at a frequency $2\pi f = \Delta\omega = (\Delta g)\mu_{B}B/\hbar$. $Z$-rotations can be turned on by shifting the detuning closer to the charge anti-crossing where $J$ is larger, driving oscillations around the equator of the Bloch sphere, Fig \ref{fig:Fig1}(c). The spin state is detected using Pauli blockade, combined with a remote charge sensor that detects whether the qubit state passed through the $(2,0)$ charge state or was blockaded in $(1,1)$ during the readout stage\cite{HarveyCollard2017}.

The spin splitting of an electron in a QD is governed by an effective Zeeman Hamiltonian of the form $H_{\mathrm{eff}} = \frac{\mu_{B}}{2}\mathbf{B} \cdot \mathbf{g} \cdot \boldsymbol{\sigma}$, where $\boldsymbol{B}$ is the magnetic field vector, $\boldsymbol{\sigma}$ is the vector of Pauli spin matrices $(\sigma_{x}, \sigma_{y}, \sigma_{z})$ and $\boldsymbol{g}$ is the electron $g$-tensor. Including the $H_{R}$ and $H_{D}$ SO Hamiltonians perturbatively leads to an effective $g$-tensor of the form
\begin{equation}
\mathbf{g} = \left( 
\begin{array}{ccc}
g_{\perp} - \frac{2}{\mu_{B}} \alpha & \frac{2}{\mu_{B}} \beta & 0 \\
\frac{2}{\mu_{B}} \beta & g_{\perp} - \frac{2}{\mu_{B}} \alpha & 0 \\
0 & 0 & g_{\parallel}
\end{array}
\right).
\end{equation}
The strength of the SO interaction is predicted to depend on applied electric field, lateral confinement, valley-orbit configuration, and the atomic-scale structure of the interface (see supplementary material). Consequently, the local interfacial and electrostatic environments particular to each QD produce differences in effective $g$-tensor, Fig.  \ref{fig:Fig2}(f). This will act as a difference in effective in-plane magnetic field, modifying the electron spin splitting between dots and drive rotations at a frequency
\begin{equation}\label{eq:angledependence}
f_{\mathrm{rot}}(\theta,\phi) = \Delta_{\mathrm{SO}}(\theta,\phi)/h =(\Delta g(\theta,\phi)) \mu_{B} B / h = \frac{2}{h}\vert\langle S\vert H\vert T_{0}\rangle\vert= \vert \mathbf{B} \vert \vert \Delta \alpha - \Delta \beta \sin(2 \phi) \vert \sin^{2} (\theta), 
\end{equation}
where $\phi$ is the field direction in-plane of the interface with respect to the $[100]$ crystallographic direction and $\theta$ is the out-of-plane angle relative to $[001]$. 

In Fig. \ref{fig:Fig2}(b), we show the singlet return signal as a function of time spent at the manipulation point in $(1,1)$ as the external magnetic field is varied along the $[1\bar{1}0]$ crystallographic direction. The observed oscillations demonstrate the ability to control coherent rotations. The rotation frequency displays a clear magnetic field dependence. In Fig. \ref{fig:Fig2}(c), we plot the SO-induced rotation frequency as a function of field for both the $[110]$ and $[1\bar{1}0]$ directions. The linear dependence on field is consistent with a $g$-factor difference between the two QDs ($f = (\Delta g)\mu_{B}B/h$), whereas the difference in the slopes indicates an angular dependence for $\Delta g$. We plot the full angular dependence of the SO interaction in Figs. \ref{fig:Fig2}(d) and \ref{fig:Fig2}(e).  Figure \ref{fig:Fig2}(d) shows the measured difference in gyromagnetic ratio between the dots, $(\Delta g)\mu_{B}/h$, as a function of the in-plane angle $\phi$ relative to the [100] crystallographic direction. Dependence on the out-of-plane angle, $\theta$, is shown in Fig. \ref{fig:Fig2}(e). Here, $\phi$ is fixed along the [110] ([1$\bar{1}$0]) direction and the measured difference in gyromagnetic ratio between the dots is plotted in blue (red) as the field is tilted out of the interface plane ($\theta$ = 0 is along the $[001]$ direction). Qualitatively, the angular dependence is consistent with a SO effect, slightly different in each QD, composed of Rashba and Dresselhaus contributions.  Enhanced interface SO effects in Si have been surmised previously for in-plane B-field dependences\cite{Yang2013, Tahan2013, Hwang2017}. We plot fits to equation \eqref{eq:angledependence} along with the data in Figs. \ref{fig:Fig2}(d) and \ref{fig:Fig2}(e). We extract relative SO parameters $\Delta \alpha$ = 1.89 MHz/T and $\Delta \beta$ = 15.7 MHz/T.  The maximum useful B-field is limited by state preparation and measurement (SPAM) errors as the $S$-$T_{-}$ splitting becomes comparable to $k_{B}T$.  The maximum rotation frequency achieved for the present electrostatic confinement was near 20 MHz for fields above 1 T along the $[1\bar{1}0]$ direction.

\begin{figure}[!h]
	\centering
	\includegraphics[width=0.4\textwidth]{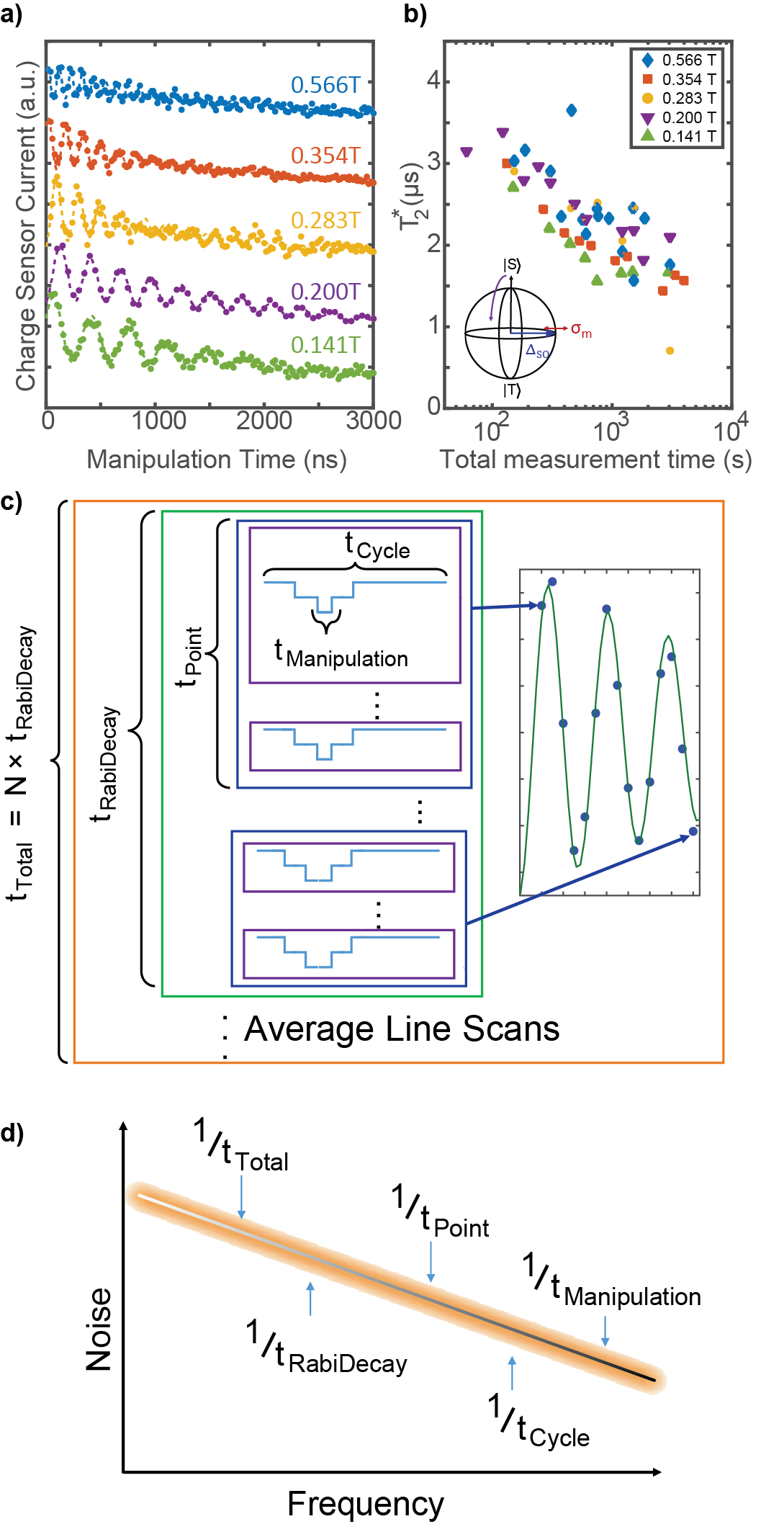}
	\caption{\textbf{Measurement time dependence}. a) Long-time (50 min) averaged measurements of singlet return signal as a function of manipulation time for several magnetic field strengths aligned along the $[1\bar{1}0]$ crystallographic direction. The data for each field has been shifted for clarity. b) The extracted $T_{2}^{*}$  as a function of total experimental measurement time. (inset) Magnetic noise creates fluctuations in the effective B-field at each QD, leading to variation in the $X$-rotation frequency throughout the measurement. c) Relevant time scales of the measurement. The shortest time scale susceptible to noise in the experiment is the time spent manipulating the qubit. In the limit of quasi-static noise, we expect the qubit to have a constant environment during this time. However, over the course of a total pulse cycle (which consists of qubit preparation and measurement and may be several ms in length), the environment may change. Furthermore, as the cycle is repeated and averaged by the lock-in for each data point, each data point is collected for a free induction decay curve. As successive curves are averaged together, the distribution of noise that is sampled grows larger. d) During the course of the measurement, the qubit is susceptible to noise in the frequency band between 1/t$_{\mathrm{Total}}$ and 1/t$_{\mathrm{Manipulation}}$.}
	\label{fig:Fig3}
\end{figure}

The ability to realize meaningful quantum information processing in MOS depends on the timescale over which environmental noise near the interface interacts with the qubit, Fig. \ref{fig:Fig3}(c,d). Although sparse, the background $^{29}\mathrm{Si}$ nuclear spins are sufficient in number to produce a slowly varying effective magnetic field, an Overhauser field.  Nuclear spin flip-flops lead to a time-variation of the Overhauser field that is quasi-static on the timescale of a single measurement instance, but can shift the rotation frequency in the time interval between measurements.  A consequence of this effect is that the decay in time of the coherent oscillations depends on the measurement integration time, as has been observed previously in ST qubits \cite{Dial2013,Eng2015}. The longer an average measurement is done, the broader the distribution of spin configurations (i.e. Overhauser fields) sampled. The ensemble-averaged singlet return signal as a function of time spent driving rotations in the $(1,1)$ region, with an external magnetic field oriented along the $[1\bar{1}0]$ crystallographic axis, is shown in Fig. \ref{fig:Fig3}(a). The decay in oscillation amplitude fits a Gaussian form consistent with quasi-static noise \cite{Dial2013}, and characteristic $T_{2}^{*}$ is extracted assuming a functional time dependence of $\exp[-(t/T_{2}^{*})^{2}]$ for the oscillation decay envelope.

In Fig \ref{fig:Fig3}(b) we examine the dependence of our results on measurement time and B-field. We find a long-averaging inhomogeneous dephasing time of $T_{2}^{*} = 1.6 \pm 0.6 \ \mathrm{\mu s}$, which is consistent with other experimental results\cite{Wu2014,Eng2015} and  theoretical estimates\cite{Assali2011, Witzel2012,Witzel2012b} (see supplementary material) of dephasing due to hyperfine coupling of the QD electron wave function with residual $^{29}\mathrm{Si}$ in the isotopically-enriched Si host. By measuring at faster timescales, an increased $T_{2}^{*} \! \sim \! 4 \ \mathrm{\mu s}$ is observed. The absence of a B-field dependence suggests that the SO coupling does not contribute appreciably to $T_{2}^{*}$. Therefore, the $T_{2}^{*}$ observed at the MOS interface is consistent with expectations of the enriched bulk Si and there is no evidence of slow noise due to the MOS interface at this enrichment level.

\begin{figure}[!h]
	\centering
	\includegraphics[width=0.95\textwidth]{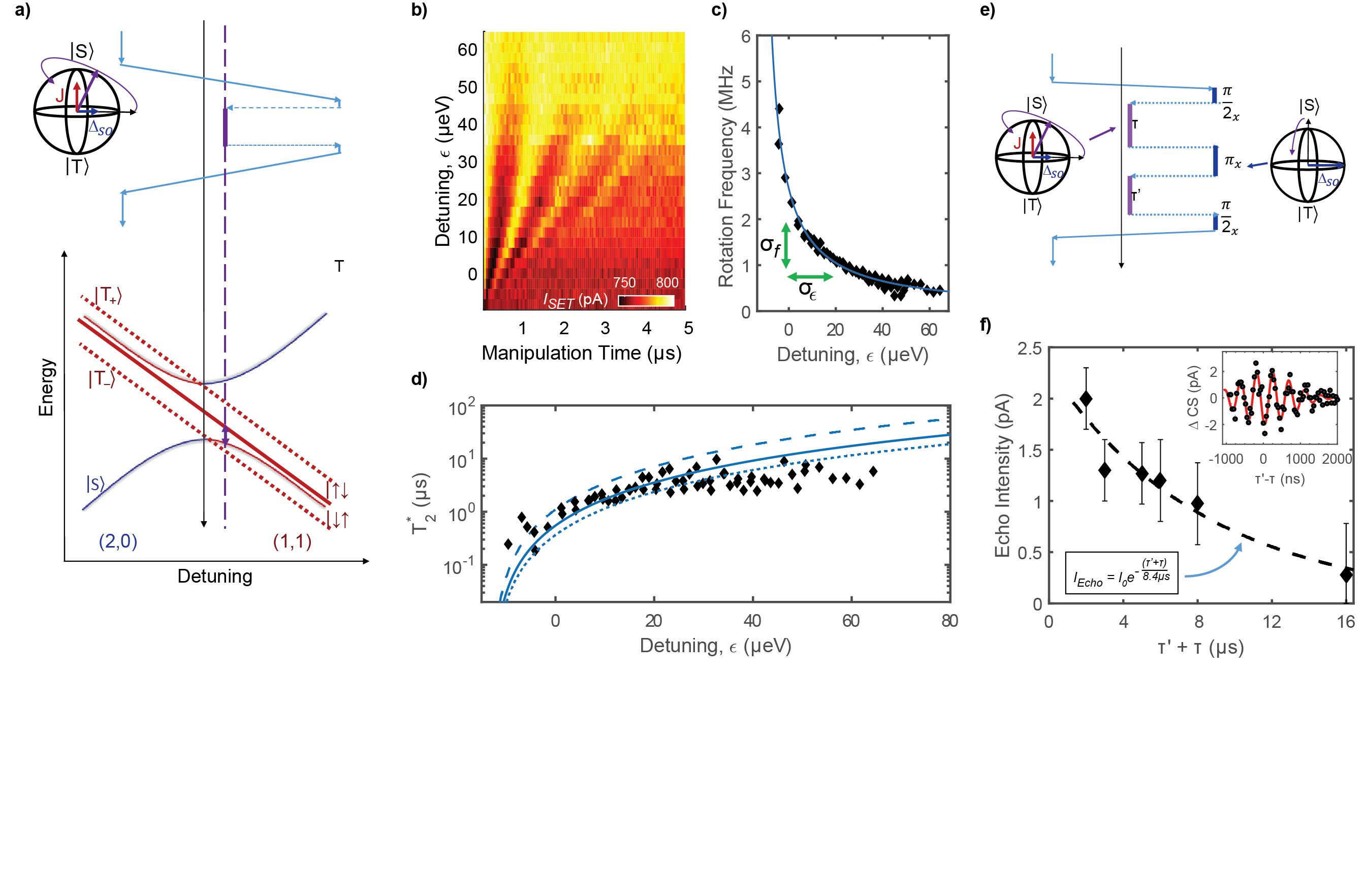}
	\caption{\textbf{$Z$ rotations and noise}. a) Energy Diagram and gate pulse schematic for controlling exchange rotations. We initialize the qubit into the S(2,0) ground state and ramp adiabatically, such that it transfers to the ground in the (1,1) charge sector. A fast pulse to and from a detuning, $\epsilon$, where $J$ is substantial drives coherent rotations around an axis depending on both $J$ and $\Delta_{\mathrm{SO}}$. Returning to the (2,0) charge sector adiabatically projects the states onto the $S(2,0)$ and $T_{0}(1,1)$ basis for measurement. b) Measured charge sensor current as a function of the time spent rotating for various detuning points. Here, high current corresponds to a higher probability of measuring a singlet. c) The extracted rotation frequency vs. detuning. The blue line is a fit to the form $\sqrt{J(\epsilon)^{2} + \Delta_{\mathrm{SO}}^{2}}$ where $J(\epsilon) \propto \epsilon^{-1}$. d) Extracted $T_{2}^{*}$  as a function of detuning. We also plot the long integration time values from Fig. 3(a). The blue lines are fits to the form $T_{2}^{*} =\frac{1}{\sqrt{2}\pi\sigma_{e}} \cdot \vert\frac{df}{d\epsilon}\vert^{-1}$, where the extracted charge noise, $\sigma_{\epsilon}$, is 1.0 $\mu$eV (dashed), 2.0 $\mu$eV (solid) and 3.0 $\mu$eV (dotted). e) Gate pulse schematic for a Hahn-echo sequence. We initialize the qubit into the S(2,0) ground state and transfer the system to the (1,1) charge sector with a fast adiabatic pulse such that it remains in a singlet state. Combinations of $\Delta_{\mathrm{SO}}$-rotations about the X-axis and $J$-rotations about a second axis provide access to entire Bloch sphere. This echo sequence counteracts low frequency noise, prolonging qubit coherence. f) Hahn-echo amplitude as a function of total time, $\tau' + \tau$, exposed to charge noise at detuning $\epsilon$. A fit to an exponential decay gives $T_{2e}^{\mathrm{echo}} = 8.4 \ \mathrm{\mu s}$. (inset) Measured echo signal for $\tau$ $=$ 1 $\mu$s with B = 0.141 T along the $[110]$ direction. The echo signal has an oscillation frequency corresponding to $\Delta_{\mathrm{SO}}$ and a Gaussian envelope around $\tau$ = $\tau'$ with a decay due to the inhomogeneous dephasing time of $T_{2e}^{*}$ $=$ 1 $\mu$s.}	
	\label{fig:Fig4}
\end{figure}

A second axis of coherent control for ST qubits is achieved through the tunable exchange coupling of the $(1,1)$ and $(2,0)$ charge states. This leads to hybridization between the $(2,0)$ and $(1,1)$ charge states and an exchange splitting, $J(\epsilon)$, between the $S$ and $T_{0}$ qubit states that depends on detuning, $\epsilon$, Fig. \ref{fig:Fig4}(a). By varying the strength of this interaction, we can achieve controlled coherent rotations, as demonstrated in Fig. \ref{fig:Fig4}(b). Here, as described in Ref. \cite{Petta2005}, we initialize into a $S(2,0)$ ground state and then adiabatically separate the electrons into the $(1,1)$ charge configuration where $J(\epsilon)$ is nearly zero and the qubit is initialized in the ground state of the SO field ($\ket{\uparrow\downarrow}$ or $\ket{\downarrow\uparrow}$), a superposition of the $S(1,1)$ and $T_{0}(1,1)$ states. We apply a fast pulse to and from finite $J(\epsilon)$ at $\epsilon$ near 0 for some waiting time, which rotates the qubit state around the Bloch sphere about a rotation axis depending on both $J$ and $\Delta_{\mathrm{SO}}$, the SO induced splitting of the $\ket{\uparrow\downarrow}$ and $\ket{\downarrow\uparrow}$ states (Fig. \ref{fig:Fig4}(a)). For this experiment, we apply a field of 0.2 T along the $[100]$ direction, which provides a small (0.5 MHz) residual $X$-rotation frequency. At detuning near $\epsilon$ = 0, we observe an increased rotation frequency, Fig \ref{fig:Fig4}(c). As the exchange pulse moves to deeper detuning, we observe a decrease in rotation frequency as well as visibility.  This is expected as $J$ decreases and the rotation axis tilts towards the direction of the SO field difference.

Figure \ref{fig:Fig4}(c) shows the observed rotation frequency as a function of detuning. The rotation frequency can be expressed as $\sqrt{J(\epsilon)^{2} + \Delta_{\mathrm{SO}}^{2}}$, since the two components add in quadrature. Indeed, we see that at deep detuning the rotation frequency saturates near 0.5 MHz, due to the SO field at this magnetic field strength and orientation. Figure \ref{fig:Fig4}(d) shows the dephasing time, $T_{2}^{*}$, associated with coherent rotations at each detuning. Here we have extracted $T_{2}^{*}$ by fitting a Gaussian decay envelope ($\exp[-(t/ T_{2}^{*})^{2}]$) to the rotations at each detuning point. Noise from charge fluctuations on the confinement gates causes deviations in the detuning point of the system, leading to dephasing of the qubit through changes in the rotation frequency. We measure shorter dephasing time near $\epsilon = 0$, which increases as we move to deeper detuning and eventually saturates at a few $\mu$s. We associate the saturation of $T_{2}^{*}$ at deeper detuning with the dominant noise mechanism transitioning from charge to magnetic noise due to residual background $^{29}\mathrm{Si}$. Following the method outlined in Ref. \cite{Dial2013}, we fit the rotation frequency to a smooth function to find the derivative, $df(\epsilon))/d\epsilon$. The ratio of $T_{2}^{*}$ to $\vert df / d\epsilon \vert^{-1}$ gives a root-mean-squared charge noise of $\sigma_{\epsilon} = 2.0 \pm 0.6 \ \mathrm{\mu eV}$. This agreement with the best reported charge noise values in GaAs/AlGaAs and Si/SiGe material systems of a few $\mu$eV \cite{Petersson2010, Shi2013,Dial2013, Wu2014} indicates that the poly-silicon MOS device structure is a comparable material system with respect to the magnitude of quasi-static charge noise in the limit of long time integration. Furthermore, successive measurements over the course of several weeks can be performed with no retuning of the device gate voltages, indicating that the MOS material system is an extremely stable qubit platform.

Improved decoherence can be achieved through dynamical decoupling (DD), which suppresses contributions from quasi-static noise through multi-rotation sequences that leverage time reversal symmetry.  A schematic for a Hahn-echo sequence to examine electrical noise is shown in Fig. \ref{fig:Fig4}(e). As seen in Fig. \ref{fig:Fig4}(f), a refocusing pulse can greatly extend the qubit coherence with a $T_{2e}^{\mathrm{echo}}$ of 8.4 $\mu$s for a detuning, $\epsilon$, where charge noise leads to $T_{2e}^{*}$ = 1 $\mu$s. This is comparable to what has been observed in GaAs/AlGaAs\cite{Dial2013} and Si/SiGe\cite{Eng2015}. Likewise, Hahn-echo techniques were able to improve decoherence from magnetic noise to a $T_{2m}^{\mathrm{echo}}$ of 70 $\mu$s (see supplementary material). These results illustrate our ability to extend coherence times through dynamical decoupling and unequivocally demonstrate full all-electrical control of the MOS spin-orbit driven ST qubit.

\section{Summary}
\label{sec:summary}

In previous implementations of ST qubits, dynamic nuclear polarization (DNP)\cite{Foletti2009,Nichol2017}  and micro-magnets\cite{Wu2014} have been used to create strong, stable difference in Zeeman splitting between two quantum dots and drive rotations. DNP produces a variation in the nuclear magnetic field between dots. However, this gradient must be actively pumped and the qubit is susceptible to fluctuations in the nuclear magnetic environment. A second approach, micro-magnet integration, has been used to produce static magnetic fields and allows operation in an enriched $^{28}\mathrm{Si}$ host environment. However, this additional fabrication complexity creates long-term integration challenges for extending to larger qubit systems (e.g. non-standard fabrication, uniformity, and layout constraints). Additionally, single nuclear spin-driven ST rotation has been demonstrated recently using phosphorus in enriched Si, which overcomes these challenges\cite{HarveyCollard2015, Rudolph2017}. However, deterministic fabrication with donors is non-trivial and remains a topic of ongoing research. In contrast to these other options, the SO driven ST qubit offers a relatively simple MOS implementation path.

The SO $X$-rotations have reached 20 MHz in our device, limited primarily by preparation and readout constraints. Though this is larger than what has been reported for a ST qubit in Si/SiGe using a micro-magnet\cite{Wu2014}, it is smaller than a number of other implementations mentioned above that have achieved 50 to 1000 MHz \cite{HarveyCollard2015, Nichol2017, Takeda2016}.  Increased drive frequency with SO coupling is likely possible through a number of avenues, including increasing vertical E-field (see supplementary material and Ref. \cite{Veldhorst2015b}), modifying the confinement potential (see SM), and by working with one of the QDs at higher occupation (since the two z-valleys at the hetero-interface are predicted to have opposite sign of the Dresselhaus strength (see SM and Refs. \cite{Veldhorst2015b,Ferdous2017,Ferdous2017b})). Single QDs have displayed a 140 MHz difference in ESR frequencies between N = 1 and N = 3 and E-field tunability\cite{Veldhorst2015b}, so drive frequencies of over 100 MHz seem realistic. This work also provides a theoretical foundation for an interface Dresselhaus and Rashba effect that avoids quantitative ambiguities due to gauge-dependence. This is relevant for devices that rely on SO effects at semiconductor interfaces, including emerging areas of research in topological quantum materials. Furthermore, future work also remains to establish how the microscopic details of the MOS interface affect the magnitudes of the Rashba and Dresselhaus terms. Considering the possibilities for improvement and the reduced complexity in fabrication, the SO driven ST qubit offers a promising new implementation for quantum information technology. 

Most significantly from this work, the SO driven ST qubit allows for a sensitive probe of noise properties at the MOS interface. The $T_{2}^{*}$ of order 1-2 $\mu$s observed in the magnetic noise dominated regime is consistent with decoherence expected from $^{29}\mathrm{Si}$ in the bulk. Charge noise magnitudes of $2.0 \pm 0.6 \ \mathrm{\mu eV}$ at $T_{\mathrm{e}} \! \sim \! 150 \  \mathrm{mK}$ are also comparable to other semiconductor systems.  Overall, the MOS interface shows no indication of increased negative effects relative to qubit operation. The opportunity to use MOS for highly sensitive spin coherent devices such as qubits has broad impact. 

\section{Methods}
\label{sec:methods}The DQD studied in this work was realized in a fully foundry-compatible, single-gate-layer, isotopically-enriched $^{28}\mathrm{Si}$ metal-oxide-semiconductor (MOS) device structure. The material stack consists of 200 nm of n poly-silicon and 35 nm of silicon-oxide on top of a silicon substrate with an isotopically enriched epitaxial layer hosting 500ppm residual $^{29}\mathrm{Si}$. The confinement and depletion gates are defined by electron beam lithography followed by selective dry etching of the poly-silicon. Phosphorus donors were implanted through a self-aligned implant window near the QD locations for alternative experiments. This was followed by an activation annealing process at 900 C. Biasing the poly-silicon gates confines a 2-dimensional electron gas into quantum dot potentials. One QD is used as a single electron transistor (SET) remote charge sensor for spin-to-charge conversion. The rest of the device is tuned such that a double quantum dot (DQD) is formed and the number of electrons in each QD is inferred from changes in current through the SET. Measurements were performed in a $^{3}\mathrm{He}$/$^{4}\mathrm{He}$ dilution refrigerator with a base temperature of around 8 mK. The effective electron temperature in the device was 150 mK. Fast RF lines we connected to cryogenic RC bias-T's on the sample board, which to allow for the application of fast gate pulses. An external magnetic field was applied using a 3-axis vector magnet. Additional information discussing the device and measurements is offered in the supplementary material and elsewhere\cite{Rochette2017}.

\section{Acknowledgements}

We would like to thank Rusko Ruskov for discussions. This work was performed, in part, at the Center for Integrated Nanotechnologies, an Office of Science User Facility operated for the U.S. Department of Energy (DOE) Office of Science by Los Alamos National Laboratory (Contract DE-AC52-06NA25396) and Sandia National Laboratories (Contract DE-NA-0003525). Sandia National Laboratories is a multimission laboratory managed and operated by National Technology and Engineering Solutions of Sandia, LLC, a wholly owned subsidiary of Honeywell International, Inc., for the DOE's National Nuclear Security Administration under contract DE-NA0003525.

\section{Author Contributions}

R.M.J, P.H.-C., and M.S.C. designed the experiments. R.M.J. performed the central measurements and analysis presented in this work. P.H.-C. performed supporting measurements on a similar device that establish repeatability of observations. N.T.J. developed the theoretical description of the results with the help of A.M.M., V.S., J.K.G., A.D.B., and W.M.W., providing critical insights. R.M.J., M.S.C., N.T.J., P.H.-C., A.M.M. and M.R. analyzed and discussed central results throughout the project. D.R.W., J.A., R.P.M., J.R.W., T.P., and M.S.C. designed the process flow, fabricated devices, and designed/characterized the $^{28}$Si material growth for this work. J.R.W. provided critical nanolithography steps. M.S.C. supervised the combined effort, including coordinating fabrication and identifying modeling needs for the experimental path. R.M.J. and M.S.C. wrote the manuscript with input from all co-authors.

\bibliography{SOSTDQD-4_bib}

\cleardoublepage

\section{Supplementary Information}

\appendix
  \subsection{Spin-orbit Coupling at the MOS Interface}
  While spin-orbit (SO) coupling in bulk silicon is weaker than in other materials commonly used for quantum dot devices, such as GaAs and InAs, an interface introduces SO coupling that may significantly influence qubit operation. Such effects have been documented recently elsewhere in the case of a single quantum dot in silicon \cite{Kawakami2014, Veldhorst2014, Veldhorst2015b, Ferdous2017, Ferdous2017b}. Here, we detail our model for the SO coupling associated with the interface. Our theoretical treatment is informed by the previous work of Refs \cite{Nestoklon2006,Nestoklon2008,Prada2011,Veldhorst2015b}.
  
  The Hamiltonian for a single electron in a silicon quantum dot in an arbitrary uniform magnetic field $\mathbf{B}$, without SO coupling included, is given by 
  \begin{eqnarray}
  H_{0} & = & H_{\mathrm{dot}} + H_{\mathrm{Zeeman}} \\ 
  & = & \frac{P_{x}^{2}}{2m_{\perp}}+\frac{P_{y}^{2}}{2m_{\perp}}+\frac{P_{z}^{2}}{2m_{\parallel}}+V(\mathbf{r}) + \frac{\mu_{B}}{2} \mathbf{B} \cdot \mathbf{g}_{0} \cdot \boldsymbol{\sigma} \nonumber,
  \end{eqnarray}
  where $\mathbf{P} = -i \hbar \nabla + e \mathbf{A}(\mathbf{r})$ is the kinetic momentum ($e>0$), $m_{\perp} = 0.19 m_{0}$ ($m_{\parallel} = 0.98 m_{0}$) is the transverse (longitudinal) effective mass, and $\mathbf{g}_{0} = \mathrm{diag}(g_{\perp}, g_{\perp}, g_{\parallel})$ is the bulk $g$-tensor for silicon. We take our coordinate system to be aligned along the Cartesian $[100]$, $[010]$, $[001]$ axes, with $[001]$ the interface normal. The potential $V(\mathbf{r})$ includes electrostatic confinement from voltages applied to gate electrodes and details of the interface potential. Atomic-scale features at the interface, the potential barrier height, and the vertical electric field dictate the valley splitting and valley content of the valley-orbital eigenstates \cite{Gamble2016}. Due to the strong vertical confinement, the low-lying valley-orbital eigenstates include contributions only from the $\pm z$ conduction band minima. As a consequence of the weak bulk SO coupling in silicon, $g_{\perp}$ and $g_{\parallel}$ are close to the vacuum $g$-factor of 2.0. In our double quantum dot device the bulk $g$-factor anisotropy, being common to both dots, does not manifest in significant measurable effects. The specific gauge choice for $\mathbf{A}$ has no influence on any physical observables, and we emphasize that any theoretical analysis must be gauge-invariant. When necessary for numerical calculations, we choose the convenient gauge $\mathbf{A}(\mathbf{r}) = \frac{1}{2} \mathbf{B} \times \mathbf{r}$.
  
  Assuming that we have found the valley-orbital eigenstates of the spin-independent part of $H_{0}$, $H_{\mathrm{dot}}$, we now treat the SO coupling as a perturbation. Following Refs \cite{Nestoklon2006,Nestoklon2008,Prada2011,Veldhorst2015b}, we take the SO interaction for an electron confined against an interface at $z=z_{i}$ to consist of both Rashba and Dresselhaus terms, $H_{R} = \gamma_{R} \delta(z-z_{i}) (P_{y} \sigma_{x} - P_{x} \sigma_{y})$ and $H_{D} = \gamma_{D} \delta(z-z_{i}) (P_{x} \sigma_{x} - P_{y} \sigma_{y})$, respectively \cite{Golub2004}. We emphasize the importance of the interface-localized $\delta$-function in these terms. As we will see, this leads to the SO coupling appearing at first order in perturbation theory, rather than second order if the SO coupling had taken the bulk form without the interface-localizing $\delta$-function. This latter property can be seen from the fact that, for bound valley-orbital eigenstates $\ket{v_{k}}$, the diagonal momentum matrix elements vanish, $\mtxelem{v_{k}}{\mathbf{P}}{v_{k}} = 0$, without approximation. This can be confirmed by applying the commutation identities $P_{x} = \frac{i m_{\perp}}{\hbar}\left[H_{0}, x \right]$, $P_{y} = \frac{i m_{\perp}}{\hbar}\left[H_{0}, y \right]$, and $P_{z} = \frac{i m_{\parallel}}{\hbar}\left[H_{0}, z \right]$. However, the interface-constrained diagonal matrix elements $\mtxelem{v_{k}}{\delta(z-z_{i})\mathbf{P}}{v_{k}}$ may be non-zero, in general, due to the cyclotron orbits established by an applied magnetic field.
  Note that, due to the crystal symmetry of silicon, a vertical shift of a pristine $(001)$ interface by $z \to z+a_{0}/4$, where $a_{0} = 0.543 \ \mathrm{nm}$ is the lattice constant, is equivalent to an in-plane rotation by an angle $\pi/2$. Consequently, while a Rashba term $P_{y} \sigma_{x} - P_{x} \sigma_{y}$ is invariant, a Dresselhaus term $P_{x} \sigma_{x} - P_{y} \sigma_{y}$ must change sign under such a transformation\cite{Nestoklon2008}, since a $\pi/2$ rotation maps $P_{y} \to P_{x}$, $P_{x} \to -P_{y}$, $\sigma_{y} \to \sigma_{x}$, and $\sigma_{x} \to -\sigma_{y}$. Hence, we assign the Dresselhaus coupling factor a dependence $\gamma_{D}(z_{i}) = \gamma_{D} \cos(4 \pi z_{i}/a_{0})$ to capture the rapidly oscillatory behavior of the sign of the Dresselhaus coupling as a function of interface position.
  
  To proceed with identifying the contributions of $H_{\mathrm{SO}}$, we must evaluate the interface-constrained momentum matrix elements $\mtxelem{v_{k}}{\delta(z-z_{i}) \mathbf{P}}{v_{j}}$ as a function of the applied magnetic field, $\mathbf{B}$. We note that the effective SO coupling strengths $\gamma_{R}$ and $\gamma_{D}$ should be expected to depend intimately on the atomistic details of the interface \cite{Nestoklon2006, Nestoklon2008, Prada2011}. For the purposes of this analysis, we wrap such details into Rashba and Dresselhaus coupling strengths $\alpha_{R}$ and $\beta_{D}$, respectively, and treat them as fit parameters. Future work will address the question of capturing the short length-scale physics of interface SO effects within a multi-valley effective mass theory framework\cite{Gamble2015}, in the spirit of previous analyses of valley splitting statistics in the presence of interface disorder\cite{Gamble2016}.
  
  The valley composition of the valley-orbital eigenstates $\ket{v_{k}}$ is dictated by the relative phase between the $+z$ and $-z$ valley components. Within a simplified envelope function picture (see e.g. Ref. \cite{Friesen2007}), the low-lying valley components are given by
  \begin{eqnarray}
  \ket{+z} & = & e^{i k_{0} z} u_{+z}(\mathbf{r}) \psi(\mathbf{r}) \\
  \ket{-z} & = & e^{-i k_{0} z} u_{-z}(\mathbf{r}) \psi(\mathbf{r}),
  \end{eqnarray}
  where $k_{0}=0.84\pi/a_{0}$ is the position of the conduction band minimum, $u_{\pm z}(\mathbf{r})$ are the lattice-commensurate Bloch functions for silicon's $\pm z$ conduction band minima, and $\psi(\mathbf{r})$ is an envelope function. The lowest two valley-orbital eigenstates are, then
  \begin{eqnarray}
  \ket{v_{0}} & = & \frac{1}{\sqrt{2}} (\ket{+z} + e^{i \phi_{\mathrm{v}}} \ket{-z}) \\
  \ket{v_{1}} & = & \frac{1}{\sqrt{2}} (\ket{+z} - e^{i \phi_{\mathrm{v}}} \ket{-z}),
  \end{eqnarray}
  where $\phi_{\mathrm{v}}$ is the valley phase factor. As mentioned previously, the value of $\phi_{\mathrm{v}}$ and the valley splitting $\Delta_{\mathrm{vs}} = \mtxelem{v_{1}}{H_{\mathrm{dot}}}{v_{1}} - \mtxelem{v_{0}}{H_{\mathrm{dot}}}{v_{0}}$ is dictated by details of the interface and associated confinement potential.
  
  In particular, we approximate 
  \begin{equation}
  \mtxelem{v_{0}}{\delta(z \! - \! z_{i})P_{j}}{v_{0}} =  \frac{c}{2}\left(1\!+\!\cos\left(\phi_{\mathrm{v}}\!-\!2k_{0}z_{i}\right)\right) \mtxelem{\psi}{\delta(z \! - \! z_{i}) P_{j}}{\psi},
  \end{equation}
  where 
  \begin{equation}
  \mtxelem{\psi}{\delta(z \! - \! z_{i}) P_{j}}{\psi} = \iint\!\!\mathrm{d}x\mathrm{d}y\ \psi^{*}(x,y,z_{i}) P_{j} \psi(x,y,z_{i}),
  \end{equation}
  with $\psi(\mathbf{r})$ the envelope function and $c$ an unknown real parameter that depends on details of the Bloch function at the interface. Similarly, for the first excited valley state we'd obtain 
  \begin{equation}
  \mtxelem{v_{1}}{\delta(z \! - \! z_{i})P_{j}}{v_{1}} = \frac{c}{2}\left(1\!-\!\cos\left(\phi_{\mathrm{v}}\!-\!2k_{0}z_{i}\right)\right) \mtxelem{\psi}{\delta(z \! - \! z_{i}) P_{j}}{\psi}.
  \end{equation}
  
\begin{figure}[!h]
  	\centering
  	\includegraphics[width=0.5\textwidth]{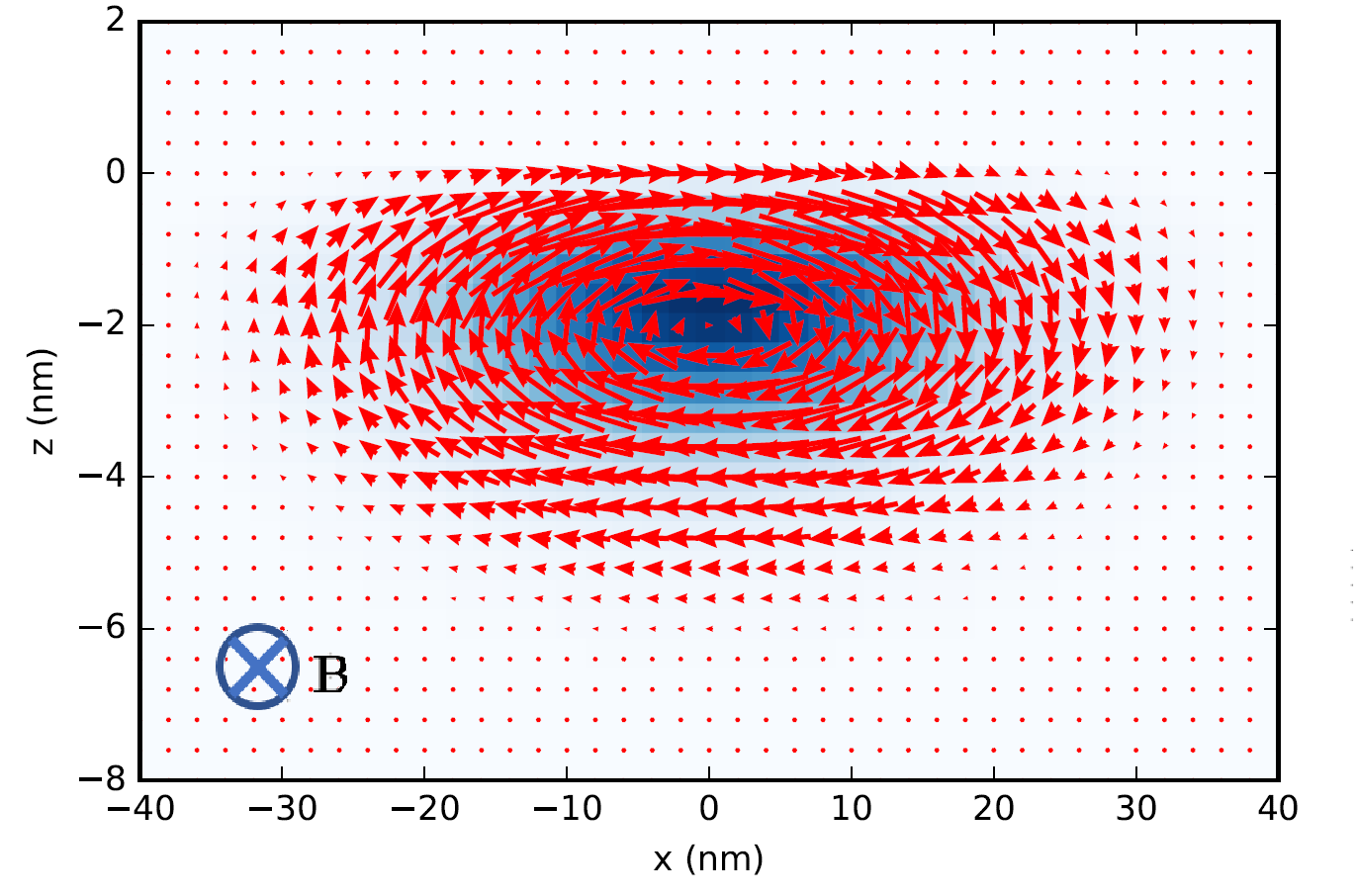}
  	\caption{\textbf{Quantum dot wave function}. Kinetic momentum density and cyclotron motion of the ground state of a quantum dot with an applied B-field $\mathbf{B} = B \hat{y}$, for a slice at $y=0$. In blue we show the probability density of the electron. The interface with offset $U_{0} = 3 \ \mathrm{eV}$ is located at $z=0$, vertical field is $F_{z} = 10 \ \mathrm{MV/m}$, and lateral confinement energies are $\hbar \omega_{x} = 1 \ \mathrm{meV}$, $\hbar \omega_{y} = 3 \ \mathrm{meV}$. Notice that $\mtxelem{\psi}{P_{x}}{\psi}=0$, while $\mtxelem{\psi}{\delta(z)P_{x}}{\psi} > 0$.}
  	\label{fig:SIMomentumDensity}
  \end{figure}

  To investigate the momentum matrix element $\mtxelem{\psi}{\delta(z \! - \! z_{i}) P_{j}}{\psi}$ with respect to the envelope function, we have implemented a (valley-free) finite-difference discretization of a Hamiltonian for a quantum dot that is harmonically confined laterally, with a uniform vertical electric field $F_{z}$ and an interface with energy offset $U_{0}$,
  \begin{eqnarray}
  H & = & \frac{P_{x}^{2}}{2m_{\perp}}+\frac{P_{y}^{2}}{2m_{\perp}}+\frac{P_{z}^{2}}{2m_{\parallel}} \\
  & & + \frac{1}{2}m_{\perp}\omega_{x}^2x^{2}+\frac{1}{2}m_{\perp}\omega_{y}^2y^{2}+F_{z}z+U_0\Theta(z) \nonumber
  \end{eqnarray}
  The harmonic confinement energies $\hbar \omega_{x}$, $\hbar \omega_{y}$ are allowed to be distinct, describing an anisotropically-shaped quantum dot. From qualitative fits to a numerical analysis, we find the following functional form for the matrix elements with respect to the ground state envelope function:
  \begin{eqnarray}
  \mtxelem{\psi}{\delta(z \! - \! z_{i}) P_{x}}{\psi} & \approx & (a-b \hbar \omega_{x}) F_{z}^{2/3} B_{y} \\
  \mtxelem{\psi}{\delta(z \! - \! z_{i}) P_{y}}{\psi} & \approx & -(a-b \hbar \omega_{y}) F_{z}^{2/3} B_{x},
  \end{eqnarray}
  where for dot confinement energies of $\mathcal{O}(\mathrm{meV})$ we find $(b \times 1 \ \mathrm{meV})/a \approx 2 \%$. The $F_{z}^{2/3}$ dependence is consistent with what is expected for a triangular vertical confinement potential \cite{Veldhorst2015b}. Notice that these matrix elements depend weakly on the lateral confinement energies, with the dominant dependence on the vertical electric field and transverse magnetic field. This qualitative functional dependence on magnetic and vertical electric fields is consistent with the analysis of Ref \cite{Veldhorst2015b}.
  In Fig. \ref{fig:SIMomentumDensity}, we plot a representative momentum density, indicating the cyclotron orbits induced by the applied magnetic field.
  
  Combining the envelope function and valley components together, we obtain the following functional form for the interface-constrained momentum matrix elements:
  \begin{eqnarray}
  \mtxelem{v_{0}}{\delta(z \! - \! z_{i})P_{x}}{v_{0}} & \propto & \left(1\!+\!\cos\left(\phi_{\mathrm{v}}\!-\!2k_{0}z_{i}\right)\right) \nonumber\\
  & & \times (a-b \hbar \omega_{x}) F_{z}^{2/3} B_{y} \nonumber\\
  & = & \lambda(\phi_{\mathrm{v}}, z_{i}, F_{z}, \hbar \omega_{x}) B_{y} \\
  \mtxelem{v_{0}}{\delta(z \! - \! z_{i})P_{y}}{v_{0}} & \propto & -\left(1\!+\!\cos\left(\phi_{\mathrm{v}}\!-\!2k_{0}z_{i}\right)\right) \nonumber\\
  & & \times (a-b \hbar \omega_{y}) F_{z}^{2/3} B_{x} \nonumber \\
  & = & -\lambda(\phi_{\mathrm{v}}, z_{i}, F_{z}, \hbar \omega_{y}) B_{x},
  \end{eqnarray}
  where $\lambda(\phi_{\mathrm{v}}, z_{i}, F_{z}, \hbar \omega)$ is a function that encodes the dependence on valley phase, interface location, vertical electric field, and lateral confinement.
  
  While the SO interaction will induce non-zero matrix elements between valley-orbital eigenstates such as $\mtxelem{v_{0} \! \uparrow}{H_{\mathrm{SO}}}{v_{1} \! \downarrow}$, the influence of these matrix elements will be suppressed by the valley splitting $\Delta_{\mathrm{vs}}$ \cite{Veldhorst2015b}. Since valley splitting in MOS systems is typically relatively large ($\mathcal{O}(100 \ \mathrm{\mu eV})$ in this experiment) and these inter-valley eigenstate matrix elements appear to second order in perturbation theory, we neglect them here.
  
  Within the subspace spanned by the tensor product of the lowest valley-orbital eigenstate $\ket{v_{0}}$ and $\sigma_{z}$ spin eigenstates $\ket{\uparrow}$, $\ket{\downarrow}$, $\lbrace \ket{v_{0} \! \uparrow}, \ket{v_{0} \! \downarrow}$, we can express the SO Hamiltonian as 
  \begin{equation}
  H_{\mathrm{SO}} = (-\gamma_{R} \lambda_{y} B_{x} + \gamma_{D} \lambda_{x} B_{y}) \sigma_{x} + (-\gamma_{R} \lambda_{x} B_{y} + \gamma_{D} \lambda_{y}) \sigma_{y}.
  \end{equation}
  If we make the approximation that the dot is nearly symmetric, $\omega_{x} \approx \omega_{y}$, then this reduces to the form
  \begin{eqnarray}
  H_{\mathrm{SO}} & = & (-\alpha_{R} B_{x} + \beta_{D} B_{y}) \sigma_{x} + (-\alpha_{R} B_{y} +\beta_{D} B_{x}) \sigma_{y} \nonumber \\
  & = & \frac{\mu_{B}}{2} \mathbf{B} \cdot \mathbf{g}_{\mathrm{SO}} \cdot \boldsymbol{\sigma},
  \end{eqnarray}
  where
  \begin{equation}
  \mathbf{g}_{\mathrm{SO}} = \frac{2}{\mu_{B}} \left(
  \begin{array}{ccc}
  -\alpha_{R} & \beta_{D} & 0 \\
  \beta_{D} & -\alpha_{R} & 0 \\
  0 & 0 & 0
  \end{array}
  \right).
  \end{equation}
  Consequently, the interface SO interaction in a quantum dot can be represented as a modified $g$-tensor $\mathbf{g}$ = $\mathbf{g}_{0} + \mathbf{g}_{\mathrm{SO}}$, where 
  \begin{equation}
  \mathbf{g} = 
  \left( \begin{array}{ccc}
  g_\perp - 2 \alpha_{R}/\mu_{B} & 2 \beta_{D}/\mu_{B} & 0 \\
  2 \beta_{D}/\mu_{B} & g_{\perp} - 2 \alpha_{R}/\mu_{B} & 0 \\
  0 & 0 & g_{\parallel}
  \end{array} \right).
  \end{equation}
  Note here that the total $g$-tensor $\mathbf{g}$ need not be symmetric, since any asymmetry in the quantum dot geometry may result in $g_{xy} \neq g_{yx}$, in general. However, in our fitting to the present experimental data we have observed satisfactory agreement when assuming a symmetric $g$-tensor. Future measurements with reduced statistical uncertainty or more anisotropic dot geometries may allow for this effect to be probed.
  
  In the regime of deep detuning, for which the two electrons in the DQD are well delocalized into the $(1,1)$ charge configuration, we can treat the interface SO coupling as producing a distinct effective $g$-tensor in the left and right dots, $\mathbf{g}_{L}$ and $\mathbf{g}_{R}$. That is, the SO Hamiltonian transforms the Zeeman Hamiltonian for the two-electron problem into
  \begin{equation}
  H_{\mathrm{Zeeman}} = \frac{\mu_{B}}{2} \mathbf{B} \cdot \mathbf{g}_{L} \cdot \boldsymbol{\sigma}_{L} + \frac{\mu_{B}}{2} \mathbf{B} \cdot \mathbf{g}_{R} \cdot \boldsymbol{\sigma}_{R},
  \end{equation}
  where $\boldsymbol{\sigma}_{L}$ ($\boldsymbol{\sigma}_{R}$) is the vector of Pauli operators acting on an electron in the left (right) quantum dot.
  
  	We now show how this $g$-tensor difference appears in terms of the basis states $\lbrace \ket{S(1,1)}, \ket{T_{+}(1,1)}, \ket{T_{0}(1,1)}, \ket{T_{-}(1,1)}\rbrace$, where we follow the convention of Ref. \cite{Stepanenko2012}:
  	
  	\begin{eqnarray}
  	\ket{S(1,1)}& = & \frac{1}{\sqrt{2}} \left( c_{L \! \uparrow}^{\dagger} c_{R \! \downarrow}^{\dagger} - c_{L \! \downarrow}^{\dagger} c_{R \! \uparrow}^{\dagger}\right) \ket{\emptyset} \\
  	\ket{T_{+}(1,1)}& = & c_{L \! \uparrow}^{\dagger} c_{R \! \uparrow}^{\dagger} \ket{\emptyset} \nonumber \\
  	\ket{T_{0}(1,1)}& = & \frac{1}{\sqrt{2}} \left( c_{L \! \uparrow}^{\dagger} c_{R \! \downarrow}^{\dagger} + c_{L \! \downarrow}^{\dagger} c_{R \! \uparrow}^{\dagger}\right) \ket{\emptyset} \nonumber \\
  	\ket{T_{-}(1,1)}& = & c_{L \! \downarrow}^{\dagger} c_{R \! \downarrow}^{\dagger} \ket{\emptyset}, \nonumber
  	\end{eqnarray}
  	where $c_{L \! \uparrow}^{\dagger}$ ($c_{R \! \uparrow}^{\dagger}$) creates an electron in the left (right) quantum dot with spin up in the eigenbasis of $\sigma_{z}$ (relative to the crystallographic axis $[001]$) and $\ket{\emptyset}$ is the zero-electron state. Given this set of basis states and defining
  	\begin{eqnarray}
  	\delta \mathbf{b} & = & \frac{\mu_{B}}{2} \mathbf{B} \cdot (\mathbf{g}_{L}-\mathbf{g}_{R})/2 \\
  	\bar{\mathbf{b}} & = &  \frac{\mu_{B}}{2} \mathbf{B} \cdot (\mathbf{g}_{L}+\mathbf{g}_{R})/2, \nonumber
  	\end{eqnarray}
  	we can now write down the Zeeman Hamiltonian incorporating SO coupling:
  	
  	\begin{equation*}
  	H_{\mathrm{Z}} = \left(
  	\begin{array}{cccc}
  	0 &  -\sqrt{2}(\delta b_{x} \! + \! i \delta b_{y}) & 2 \delta b_{z}  & \sqrt{2}(\delta b_{x} \! - \! i \delta b_{y}) \\
  	\cdot & 2 \bar{b}_{z} & \sqrt{2}(\bar{b}_{x} \! - \! i \bar{b}_{y}) & 0 \\
  	\cdot & \cdot & 0 & \sqrt{2}(\bar{b}_{x} \! - \! i \bar{b}_{y}) \\
  	\cdot & \cdot & \cdot & -2 \bar{b}_{z}
  	\end{array}
  	\right),
  	\end{equation*}
  	where
  	\begin{eqnarray}
  	\delta b_{x} & = & \frac{1}{2} (-B_{x} \Delta \alpha + B_{y} \Delta \beta) \\
  	\delta b_{y} & = & \frac{1}{2} (-B_{y} \Delta \alpha + B_{x} \Delta \beta) \\
  	\delta b_{z} & = 0.
  	\end{eqnarray}
  	We now evaluate the unpolarized triplet spin eigenstate $\ket{\tilde{T}_{0}(1,1)}$ relative to the quantization axis dictated by the applied magnetic field, $\mathbf{B}$. Using the fact that the $g$-tensor is only weakly perturbed from its bulk value, $\vert \mathbf{g} - 2 I \vert \ll 1$, and diagonalizing the 3x3 triplet block, we obtain
  	\begin{equation}
  	\ket{\tilde{T}_{0}} = \cos \theta \ket{T_{0}} + \frac{1}{\sqrt{2}} \sin \theta (e^{i \phi} \ket{T_{-}} - e^{-i \phi} \ket{T_{+}}),
  	\end{equation}
  	where the applied B-field is taken to be
  	\begin{equation} 
  	\mathbf{B} = \vert \mathbf{B} \vert (\sin \theta \cos \phi, \sin \theta \sin \phi, \cos \theta))
  	\end{equation}
  	with respect to the crystallographic axes $[100]$, $[010]$, and $[001]$.
  	
  	Finally, to evaluate the frequency of $S/T_{0}$ rotations generated by such a difference in $g$-tensors, we need to evaluate the matrix element $\mtxelem{S}{H_{\mathrm{SO}}}{\tilde{T}_{0}}$, where $\ket{S}$ and $\ket{\tilde{T}_{0}}$ are the singlet and unpolarized triplet states defined with respect to the spin basis of the applied uniform B-field $\mathbf{B}$ defined above. We find that this rotation frequency is
  	\begin{eqnarray}
  	f_{\mathrm{rot}} & = & \frac{2}{h} \vert \mtxelem{S}{H_{\mathrm{SO}}}{\tilde{T}_{0}} \vert \\
  	& = & \frac{4}{h} \vert \sin \theta \big( \cos \phi \delta b_{x} + 2 \sin \phi \delta b_{y} \big) \vert \\
  	& = & \frac{2}{h} \vert \mathbf{B} \vert \vert \Delta \alpha - \Delta \beta \sin(2 \phi) \vert \sin^{2} \theta.
  	\end{eqnarray}
  	From the above expression, it's clear that applying a B-field normal to the interface $\theta=0$ will generate no effective magnetic field gradient. For an in-plane field, depending on the relative sign of the Rashba and Dresselhaus differences $\Delta \alpha$ and $\Delta \beta$, there will be an azimuthal angle $\phi$ that maximizes the generated $S/T_{0}$ rotation frequency.
  	If $\mathrm{sign}(\Delta \alpha) = \mathrm{sign}(\Delta \beta)$ ($\mathrm{sign}(\Delta \alpha) \neq \mathrm{sign}(\Delta \beta)$), the maximum rotation frequency will be obtained for $\phi = -\pi/4$ ($\phi = \pi/4$), i.e. B-field oriented along $[1 \bar{1} 0]$ ($[1 1 0]$).
  	Conversely, for $\vert \Delta \beta \vert \gg \vert \Delta \alpha \vert$ the minimum rotation frequency would be obtained for $\phi \approx 0$ or $\pi/2$, i.e. nearly aligned along the $[1 0 0 ]$ or $[010]$ Cartesian axes. In our experiment, with $\vert \Delta \beta / \Delta \alpha \vert \approx 8.3$, the minimum frequency should be attained with a B-field about 3.5 degrees away from the $[100]$ orientation.

  \subsection{Device Fabrication, Structure, Operation}
  
  \subsubsection{Device Structure}
  
  The singlet-triplet (ST) qubit studied in this work was fabricated in a fully foundry-compatible process using a single-gate-layer, metal-oxide-semiconductor (MOS) poly-silicon gate stack with an epitaxially-enriched $^{28}$Si epi-layer with 500ppm residual $^{29}$Si. Hall bars from the same sample wafer with the same gate oxide were used to extract the critical density ($n_{c}$ = 5.7x10$^{11}$/cm$^2$), the peak mobility ($\mu$ =4500 cm$^2$/Vs), threshold voltage ($V_{th}$ = 1.1 V), the RMS interface roughness ($\Delta$ = 2.4 $\AA{}$), and roughness correlation length ($\lambda$ = 26 $\AA{}$). An SEM image of a device fabricated nominally identically to the one used in this work and a schematic of the gate stack are shown in Fig. S\ref{fig:Supp1}(a,b). The device is operated in an enhancement mode using voltage biasing of the highly doped n+ poly-silicon gates to confine electrons to quantum dot (QD) potentials under gates LCP and UCP. The gates ULG, URG, LLG and LRG overlap with n+ regions and ohmic contacts and are biased to accumulate a two-dimensional electron gas (2DEG) under each gate. The 2DEGs act as source and drain electron reservoirs for the quantum dots. The lower half of the device is tuned such that a double quantum dot (DQD) is formed. One QD is tunnel coupled to the reservoir under LRG and the other quantum dot can only be occupied by electron tunneling through the first QD. The upper half of the device is used as a single electron transistor (SET) remote charge sensor. The SET is biased with 70 $\mu V$ (rms) AC bias at 0V DC and the current is measured with an AC lock-in technique at 979 Hz. The electron temperature, $T_{\mathrm{e}} \sim$ 150 mK, was measured by QD charge transition line width. More details about fabrication can be located in Ref \cite{Rochette2017}.

\subsubsection{Dot Occupation and Location}

\begin{figure}[!h]
	\centering
	\includegraphics[width=0.8\textwidth]{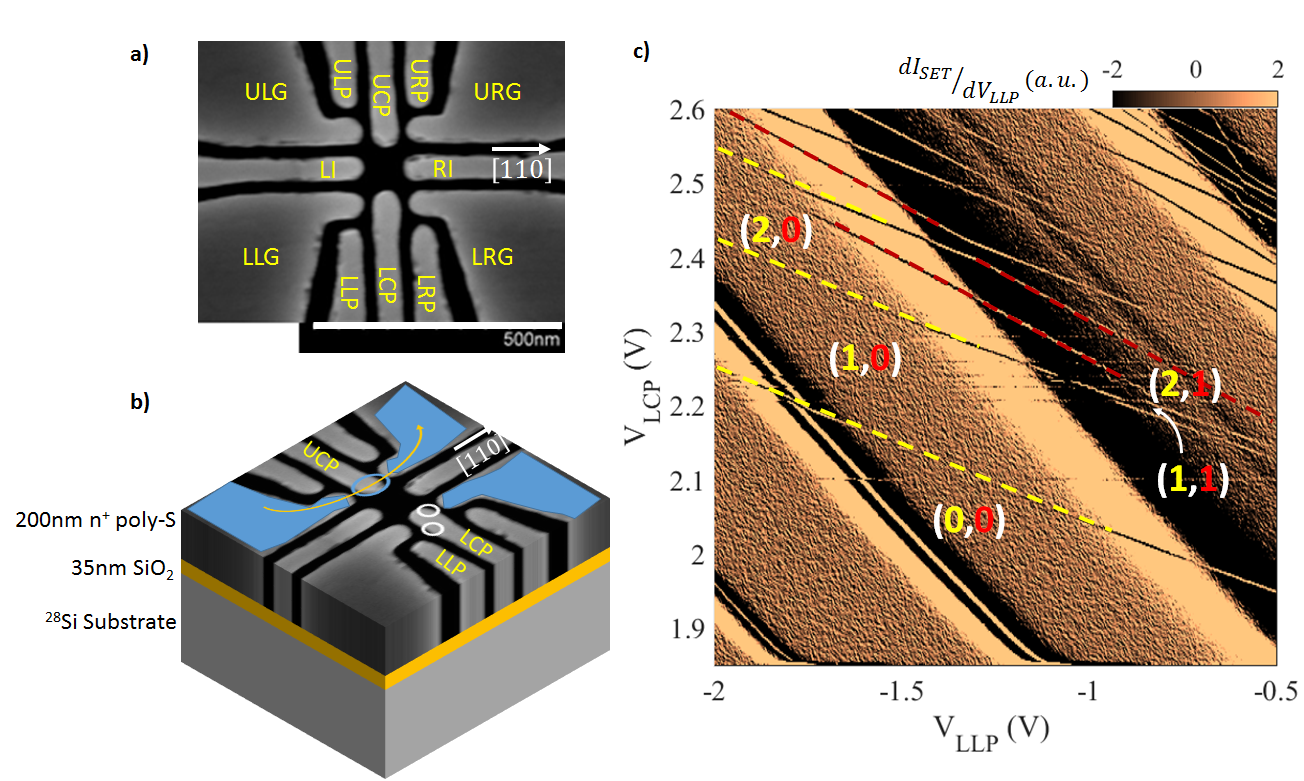}
	\caption{\textbf{Device Structure}. a) A top-down SEM of the single-layer poly-silicon gate design. The gates are labeled in yellow and the [110] crystallographic direction is indicated in white. b) A cartoon schematic of the MOS gate-stack. The 2DEG reservoirs used in these experiments are illustrated in blue with the current through the top QD SET charge sensor depicted by the yellow arrow. The approximate locations of the two QDs are represented by the white circles. c) A charge stability diagram of the DQD. Here, the gradient of the current running through the SET charge sensor is plotted as the gates LLP and LCP are varied. The broad diagonal background features are due to Coulomb peaks of the SET charge sensor. The sharp features correspond to charge transitions of objects in the lower half of the device. $QD_{1}$ (the QD closer to the electron reservoir under gate LRG) is indicated by the successive yellow dashed lines, and $QD_{2}$ is indicated by the dashed red lines. The regions in gate space corresponding to different DQD charge occupations are labeled in parentheses with the occupation of $QD_{1}$ in yellow and $QD_{2}$ in red.}
	\label{fig:Supp1}
\end{figure}

The number of electrons in each QD may be inferred from changes in current through the SET as depicted in Fig. S\ref{fig:Supp1}(c). The collection of yellow parallel lines is assigned to a QD connected to the electron reservoir under LRG, which we call QD$_{1}$. Counting from the left, we can identify the QD$_{1}$ N=1 $\rightarrow$ N=2 charging transition. A second object is observed anti-crossing with QD$_1$, which we label as the N=0 $\rightarrow$ N=1 charge transition for a second QD, QD$_{2}$. A second line is observed belonging to QD$_{2}$ in the scan, though disorder in the system makes identifying higher occupation lines difficult.  However, the presence of Pauli spin blockade at the QD$_{1}$-QD$_{2}$, (2,0)-(1,1) anti-crossing  identifies the system as a useful DQD for a ST qubit architecture (see Fig. S\ref{fig:Supp2}). To determine the locations of QD$_{1}$ and QD$_{2}$ we can use their capacitances to the nearby poly-silicon gates. By scanning combinations of the poly-silicon gates (as seen in Fig. S\ref{fig:Supp1}(c) for LLP and LCP), we can obtain the relative capacitance of both QDs to each gate compared to the capacitance of LCP, which has the strongest capacitive coupling to both QDs. We have tabulated the relative capacitances in Table \ref{table1}. 
\begin{table}[h!]
	\centering
	\caption{Gate capacitance to QDs relative to LCP}
	\label{table1}
	\begin{tabular}{lllllll}
		\hline
		\multicolumn{1}{|l||}{c$_i$/c$_{\mathrm{LCP}}$} & \multicolumn{1}{l|}{LLP} & \multicolumn{1}{l|}{LRP} & \multicolumn{1}{l|}{LI} & \multicolumn{1}{l|}{RI} & \multicolumn{1}{l|}{LLG} & \multicolumn{1}{l|}{LRG} \\ \hline \hline
		\multicolumn{1}{|l||}{QD$_{1}$} & \multicolumn{1}{l|}{0.18} & \multicolumn{1}{l|}{0.33} & \multicolumn{1}{l|}{0.24} & \multicolumn{1}{l|}{0.16} & \multicolumn{1}{l|}{0.29} & \multicolumn{1}{l|}{0.35} \\ \hline
		\multicolumn{1}{|l||}{QD$_{2}$} & \multicolumn{1}{l|}{0.27} & \multicolumn{1}{l|}{0.22} & \multicolumn{1}{l|}{0.13} & \multicolumn{1}{l|}{0.14} & \multicolumn{1}{l|}{0.29} & \multicolumn{1}{l|}{0.10} \\ \hline
		&                       &                       &                       &                      
	\end{tabular}
\end{table}
These values allow for triangulation of the dot locations, which we have indicated in S\ref{fig:Supp1}(b) with open circles. We differentiate QD$_{2}$ from an implanted donor through several observations: (1) no hyperfine component in the rotation frequency, (2) the lack of rotations at 0T B-field, (3) the ramp rates required for adiabatic transfer through the spin gap are slower than what is expected for a donor, and (4) the presence of additional lines corresponding to the QD. We find that this layout systematically produces objects near the central QD with these capacitances when the gates opposite the electron reservoir 2DEG are at low biases. For experiments investigating single QDs, these voltage potential minima may be emptied with more negative voltages on LLP or flooded by accumulating a larger 2DEG under LLG with more positive voltages.

\subsubsection{Qubit Initialization, Operation and Readout}

\begin{figure}[!h]
	\centering
	\includegraphics[width=0.8\textwidth]{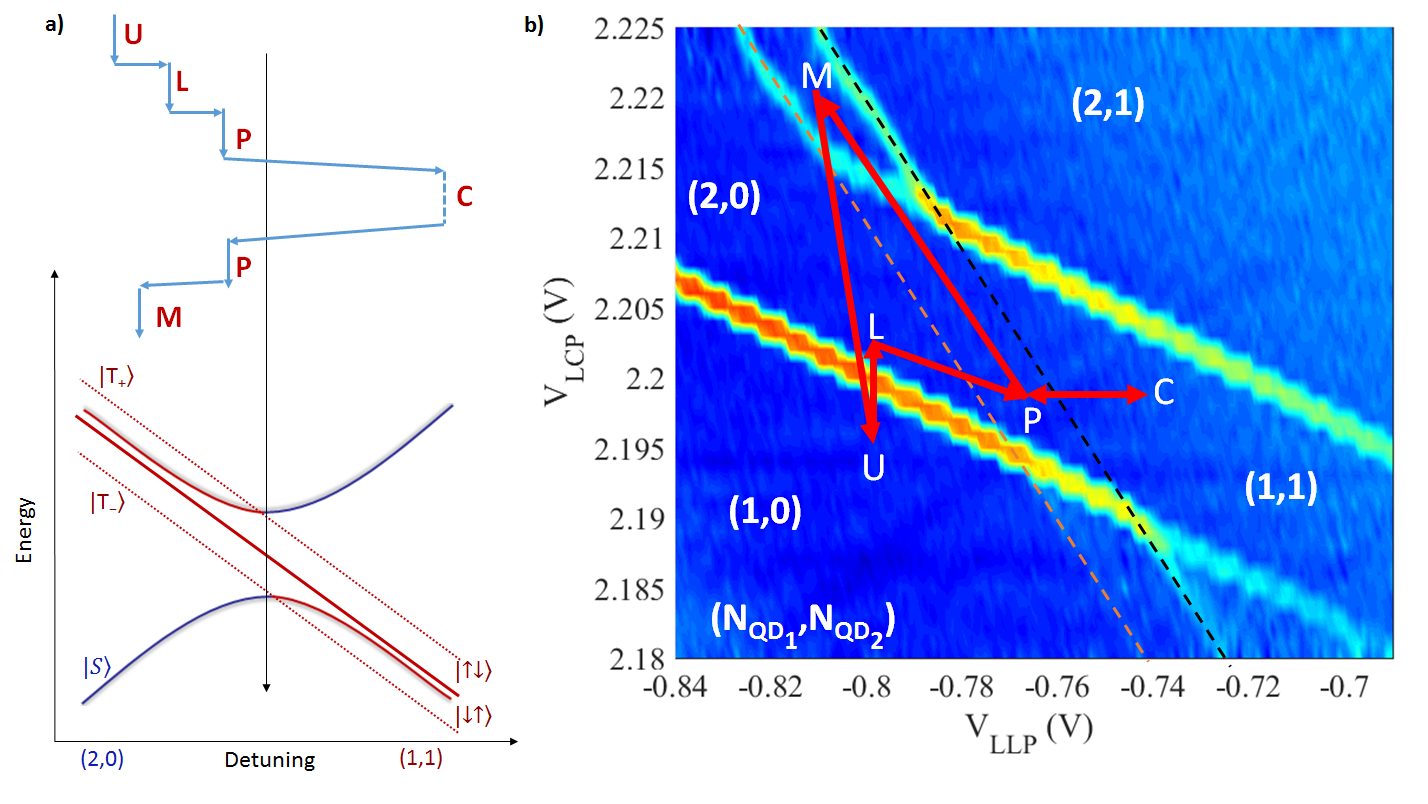}
	\caption{\textbf{Qubit Operation}. a) Energy diagram and gate pulse schematic for qubit operations. b) A pulsed charge stability diagram for the (2,0) $\rightarrow$  (1,1) anticrossing, showing the gradient of the charge sensor current. The red arrows depict a general pulse sequence for controlling the qubit, where point C may consist of several detuning pulses for different qubit manipulation sequences. The black and orange dashed lines correspond to the location of the singlet and triplet state charge preserving lines, respectively. We do not observe a change in charge sensor current at the charge preserving lines due to the orientation of the DQD dipole. }
	\label{fig:Supp2}
\end{figure}

We operate this system near the (2,0) $\rightarrow$ (1,1) spin-blockaded $(\mathrm{N_{QD_{1}}}, \mathrm{N_{QD_{2}}})$ charge anti-crossing. An energy diagram for the two-electron system is shown in Fig. S\ref{fig:Supp2}(b). The ground state charge configuration is determined by the detuning between dots, $\epsilon$, which is controlled by tuning the voltages on gates LLP and LCP. These gates are connected to cryogenic RC bias-T’s which allow the application of fast gate pulses. A schematic of the cyclical pulse sequence is shown in Fig. S\ref{fig:Supp2}(a), which is repeated as the current through the SET is monitored by the averaged AC lock-in measurements. The system is initialized in the (2,0) charge sector by first unloading (point U) the DQD into the (1,0) charge configuration and then applying an energy-selective pulse into the (2,0) charge state between the singlet and triplet energy levels such that a (2,0)S ground state is loaded (point L). The system is then plunged (point P) to a detuning ($\epsilon$ $<$  0) close to the charge anti-crossing. The electrons are then separated (point C) and qubit manipulations are performed in the (1,1) charge region ($\epsilon$ $>$ 0). The system is then pulsed back to the (2,0) charge sector (point P) where, due to Pauli spin blockade, a singlet spin state is allowed to transfer to the (2,0) charge state, but a triplet spin state is energetically blocked and remains as a (1,1) charge state \cite{Petta2005b}. We then use an enhanced latching mechanism for a spin-to-charge conversion (point M). This technique is presented in great detail in \cite{HarveyCollard2017}. There are several advantages for the use of this method. First, since an electron on the QD$_{2}$ needs to tunnel through the QD$_{1}$ to access an electron reservoir, the metastable latching state can be long lived. This allows for the measurement step in our cyclical pulse sequence to be long, compared to other points in the sequence, and dominate the time average. Second, in this approach, the charge-sensed signal differentiates between a (2,1) and a (2,0) charge state. In other words, the difference in measured current between a singlet and triplet state is the capacitive effect of adding an electron to QD$_{2}$. Thus, it does not rely on the dipole orientation of the DQD, as in traditional Pauli-blockade measurement techniques. In our case, the DQD is oriented in such a way that differentiating a (2,0) and (1,1) charge state is exceptionally difficult (observe the lack of visible inter-dot transition line in Fig. S\ref{fig:Supp2}(b)) and this method is necessary.

\subsection{Analysis Note on Extracted Data}

\subsubsection{Qubit Rotation Frequency}

\begin{figure}[!h]
	\centering
	\includegraphics[width=0.6\textwidth]{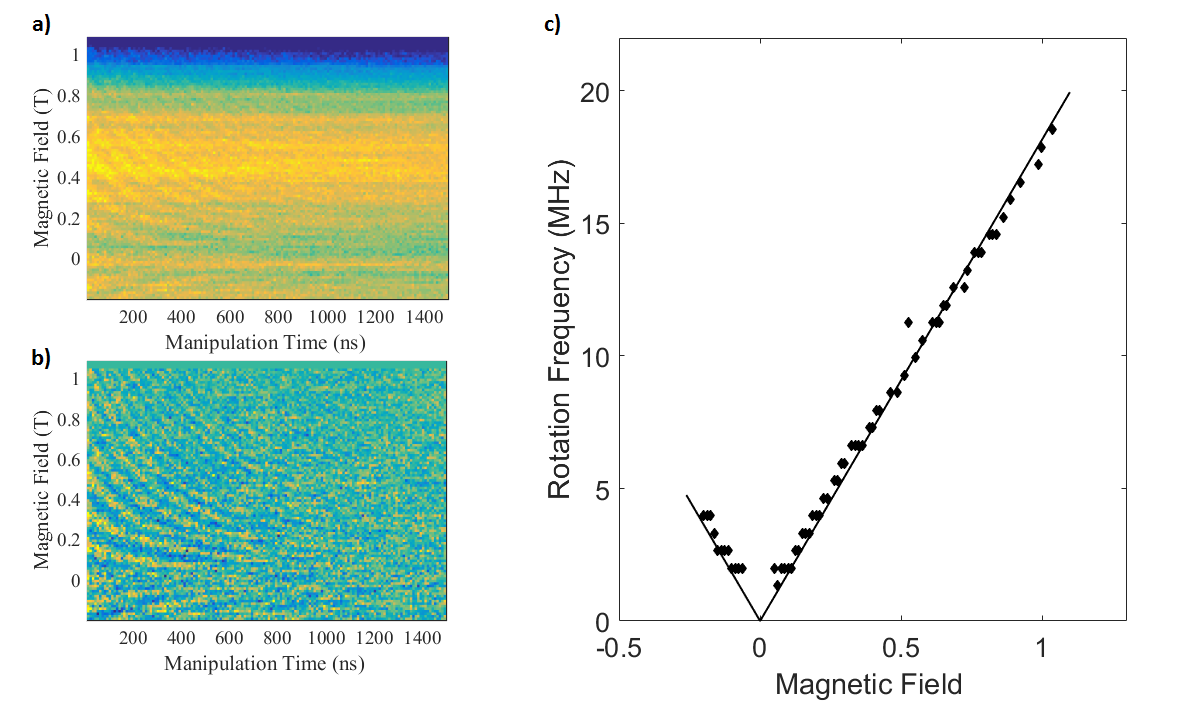}
	\caption{\textbf{Qubit Rotation Frequency Scan}. a) Measured charge sensor current as a function of manipulation time as magnetic field is stepped. b) The measured data after subtracting the linear charge sensor background current. c) The extracted rotation frequency as a function of magnetic field. The solid line is a linear fit to magnetic field strength.}
	\label{fig:Supp3}
\end{figure}

Figure S\ref{fig:Supp3}(a) shows the singlet return signal as a function of time spent at the manipulation point in $(1,1)$ as the strength of the external magnetic field is varied along the $[1\bar{1}0]$ crystallographic direction up to 1.2 T. We see that, at high magnetic field, the oscillations are difficult to observe, since the Coulomb blockade peak used for charge sensing drifts as a function of magnetic field. The qubit rotation frequency at each field was found by fitting each line scan to a Gaussian decay of the form 
\begin{equation}\label{eq:decayfit}\
I_{SET} = A \sin(2 \pi f t + \phi_{0})\exp[-(t/T_{2}^{*})^{2}] + Bt+C
\end{equation}
where all parameters are free. To help with the visualization, we subtract the background linear portion to our charge sensor signal ($Bt + C$, above), as shown in Fig. S\ref{fig:Supp3}(b).The background slope in charge sensor current is due to imperfectly separating the two electrons, such that, for some fraction of the experiments, an electron diabatically transitions through the ani-crossing, thus inelastically transferring between S(2,0) and S(1,1) on the time scale of a few $\mu$s. The rotation frequency, $f$, corresponding to the data in Fig. S\ref{fig:Supp3}(a) is plotted as a function of magnetic field in Fig. S\ref{fig:Supp3}c, indicating a 20 MHz rotation frequency at the maximum field. As can be seen in Fig. S\ref{fig:Supp2}(c), there are outlier points, which occur when a fit to equation \eqref{eq:decayfit}, produces an unphysical periodic component.
Similar techniques were used to analyze the data in the main text. The data presented in Figs. 2(c,e) of the main text were also obtained from magnetic field scan experiments and similar behavior was observed. A majority of the data fits well, and clear magnetic field strength and angular dependencies may be extracted. The data presented in Fig. 2(d), was taken with repeated scans at a given field strength and orientation. For Fig. 2(b), the linear portion of the background charge current sensor was subtracted, to clearly show the oscillations. This was useful, as slow timescale changes in the current through the charge sensor obscured the visualization.

\subsubsection{Charge Noise Characterization}

Here we describe the procedure to extract charge noise, following Ref \cite{Dial2013}. Several results in ST qubits have shown that a dominant source of dephasing during exchange oscillations can be modeled as gate-referred, quasi-static voltage fluctuations on nearby gates\cite{Dial2013,Eng2015}. These voltage fluctuations affect the energy detuning between dots, materializing as fluctuations in the exchange energy, $J$. At a given detuning, the qubit will rotate at a frequency about the Bloch sphere
\begin{equation}
f(\epsilon) = \frac{1}{h} \sqrt{J(\epsilon)^{2} + \Delta_{\mathrm{SO}}^{2}}.
\end{equation}
Therefore, we expect noise in detuning $\delta \epsilon$ to create noise in the rotation frequency $ \delta  f \sim \delta \epsilon \cdot df/d\epsilon$. For charge noise that is quasi-static, we expect a Gaussian decay of the oscillations of the form $\exp[-(t/T_{2}^{*})^2 ]$, where $T_{2}^{*}$ is the inhomogeneous dephasing time which is related to the root-mean-squared charge noise by  
\begin{equation}
\sigma_{\epsilon}=  \frac{1}{\sqrt{2}\pi T_{2}^{*}}\cdot\vert df / d\epsilon \vert^{-1}.  \end{equation}
$T_{2}^{*}$ is found for each detuning point by fitting the oscillations to Gaussian decay as shown in Fig. 4(a) of the main text. A functional form of $f(\epsilon$) is found by fitting the data in Fig. 4(c) to a smooth function. We approximate the exchange energy as $J(\epsilon) \approx t_{c}^{2}/4\epsilon$,where $J(\epsilon) = t_c^2 / 4 \epsilon$, where $t_c$ is the full-gap, inter-dot tunnel coupling, and find a good fit to $f = \frac{1}{h} \sqrt{J(\epsilon)^{2} + \Delta_{\mathrm{SO}}^{2}}$. We extract $t_c$ $=$ 0.7 $\mu$eV from the fit. From the ratio of $T_{2}^{*}(\epsilon)$ to $\vert df / d\epsilon \vert^{-1}$, for detunings less than 30 $\mu$eV, a charge noise figure of $\sigma_{e}$ = 2.0 $\pm$ 0.6 $\mu$eV can be extracted. This value agrees with reported charge noise numbers of a few $\mu$eV\cite{ Petersson2010, Shi2013,Dial2013,Wu2014}, indicating that proximity to the MOS interface does not degrade the qubit.

\subsubsection{$T_{2m}^{*}$ Magnitude}

Several theoretical estimates of $T_{2}^{*}$ in isotopically enriched silicon have been presented in the literature\cite{Assali2011,Witzel2012, Witzel2012b}. The estimate by Assali et. al. gives a $T_{2}^{*}$ of 4.4 $\mu$s for the corresponding isotopic enrichment used in our experiments (500 ppm). Note: We have included a factor of 2 because the calculations in in Ref. \cite{Assali2011} do not account for $I = \frac{1}{2}$ of the $^{29}$Si nuclei.  Witzel et. al., on the other hand, predict a $T_{2}^{*}$ of a few tens of $\mu$s, though they use a substantially larger QD radius. Following the central limit theorem, we expect $T_{2}^{*} \sim \sqrt{N_{S}}$, where $\sqrt{N_{S}}$ is the number of spinful nuclei within the QD wavefunction, and that a decrease in QD size will lead to a decrease in the inhomogeneous dephasing time.

Furthermore, these reports consider single quantum dots.  We are concerned with a DQD, in which each QD has a separate distribution of nuclear spins and the changes in the difference in hyperfine fields between QDs leads to the ST dephasing. Therefore, $T_{2}^{*}$ is inversely proportional to the amount of fluctuations in the surrounding hyperfine field. If we say that each quantum dot has a normal distribution of hyperfine fields of the form
\begin{equation}
P = \frac{e^{-(x-\mu)^{2}/(\sigma_{QD}^{2})}}{\sigma_{QD}^{2}\sqrt{2\pi}},
\end{equation}
where $\sigma_{QD}$ is the variance in hyperfine field and $\mu$ is the average hyperfine field, then the distribution of the difference in hyperfine field between the two dots is given by, 

\begin{equation}
P_{QD_{1}-QD_{2}} = \int_{-\infty}^{\infty}\int_{-\infty}^{\infty}\frac{e^{-x^{2}/(\sigma_{QD_{1}}^{2})}}{\sigma_{QD_{1}}^{2}\sqrt{2\pi}}\frac{e^{-x^{2}/(\sigma_{QD_{2}}^{2})}}{\sigma_{QD_{2}}^{2}\sqrt{2\pi}}\delta((x-y)-u)dxdy = \frac{e^{-\vert u-(\mu_{QD_{1}}-\mu_{QD_{2}})^{2}/[(\sigma_{QD_{1}}^{2})+\sigma_{QD_{2}}^{2}]}}{\sqrt{2\pi (\sigma_{QD_{1}}^{2})+\sigma_{QD_{2}}^{2})}}.
\end{equation}

This is a normal distribution with a variance of $\sqrt{\sigma_{QD_{1}}^{2}+\sigma_{QD_{2}}^{2}}$ and an average difference in Hyperfine field of $(\mu_{QD_{1}}-\mu_{QD_{2}})$. Thus, we expect 
\begin{equation}
T_{2,DQD}^{*} = T_{2,QD}^{*}/ \sqrt{2},
\end{equation}
assuming similar sized QDs.  
Taking into account the differences in QD size and the effect of two QDs, which both imply a reduction in $T_{2}^{*}$, our measured value of 1.6 $\mu$s fits well with these order of magnitude estimates. 

\subsubsection{Measurement Time Dependence of $T_{2m}^{*}$}

\begin{figure}[!h]
	\centering
	\includegraphics[width=0.65\textwidth]{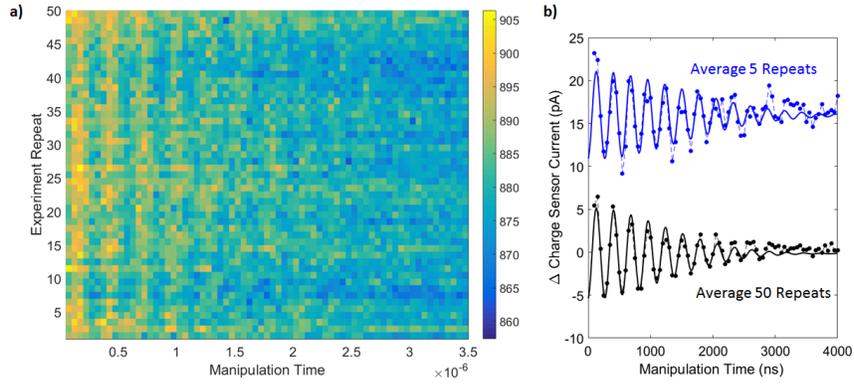}
	\caption{\textbf{Measurement Time Dependence}. a) A repeated scan of charge sensor current vs. manipulation time for a magnetic field of 0.20 T along the $[1\bar{1}0]$ crystallographic direction. b) Singlet-triplet rotation decay plots. In black, all 50 scans are averaged and a fit to the data gives a $T_2^*$ of 1.66 $\mu$s. When only 5 line scans are averaged (blue curve, shifted by 17 pA for clarity) a $T_2^*$ of 2.09 $\mu$s is extracted.}
	\label{fig:Supp4}
\end{figure}

Figure 3(b) of the main text displays a clear dependence of $T_{2}^{*}$ on the total experimental measurement time. This effect has been reported previously\cite{Dial2013,Eng2015}, and is due to the time dynamics of the random hyperfine field from the residual 500 ppm $^{29}$Si in the isotopically purified silicon host. Fluctuations in the polarization of the nuclei lead to varying magnetic fields at each QD between each experiment cycle. This leads to a varying qubit rotation frequency in each experimental cycle, which, as they are averaged together, lead to a decay in oscillation amplitude. As the experiment is measured for longer times, a larger sample of random nuclei polarization configurations, and a correspondingly larger distribution of qubit rotation frequencies is sampled. 

To obtain the plot in Figure 3(b), we repeat a measurement of charge sensor current versus manipulation time many times. An example of such a plot is shown in Fig. S\ref{fig:Supp4}(a) for a magnetic field of 0.2 T along the [110] direction. By averaging various numbers of line scans together, we can examine the effect of measurement time on $T_{2}^{*}$. Here, the total measurement time is the time for one experimental line scan times the number of scans averaged, and $T_{2}^{*}$ is extracted by fitting the envelope of the averaged data to a Gaussian decay ($\exp[-(t/T_{2}^{*})^{2}]$). Examples of the averaged data for averaging 5 and 50 line scans are shown in Fig. S\ref{fig:Supp4}(b).

\subsubsection{Hahn-Echo Measurements}

\begin{figure}[!h]
	\centering
	\includegraphics[width=0.9\textwidth]{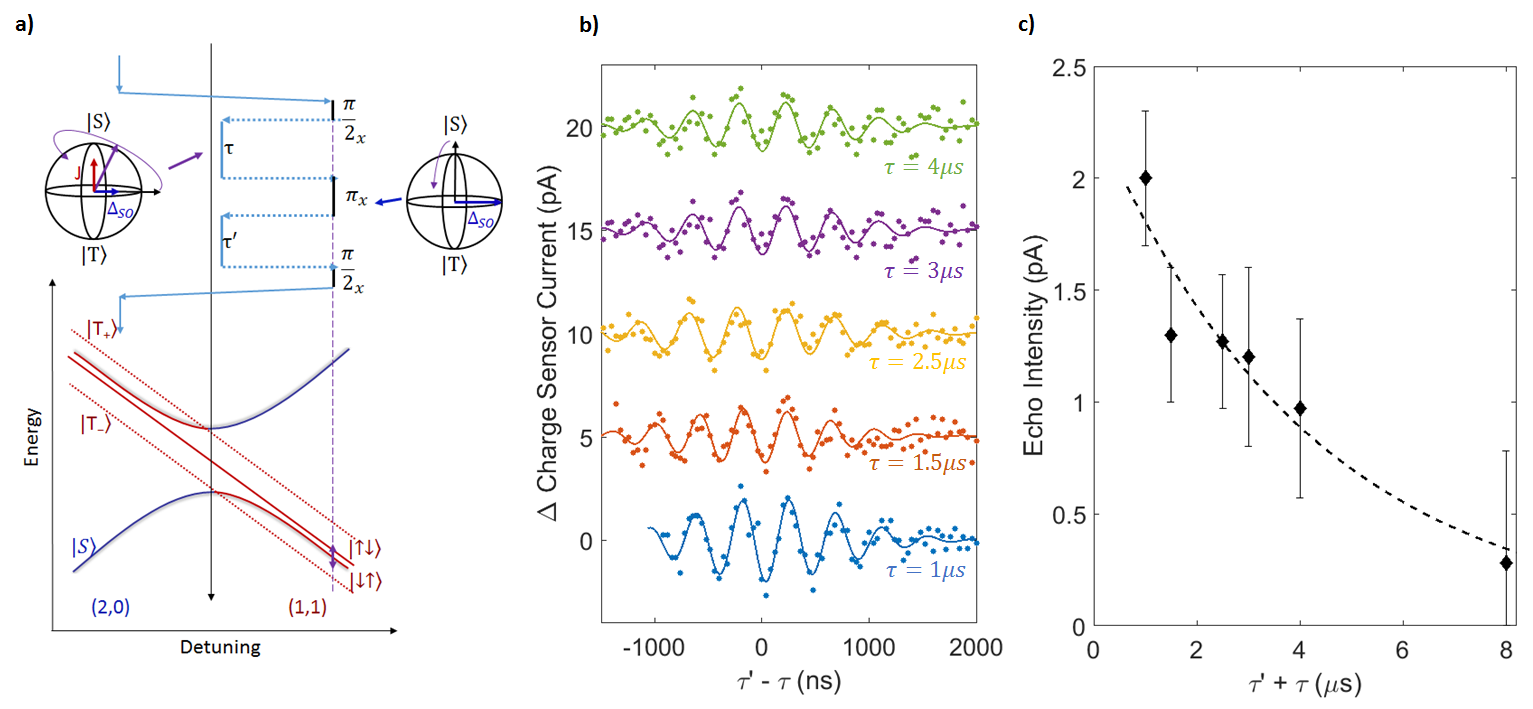}
	\caption{\textbf{Charge Noise Hahn Echo}. a) The qubit is initialized in the S(2,0) ground state and a fast adiabatic pulse transfers the system to the (1,1) charge sector such that it remains in a singlet state. The state is allowed to evolve for some time corresponding to  $\pi/2$ pulse about the X-axis and rotates the spin state to the equator of the Bloch sphere. A pulse to a detuning, $\epsilon$, where $J$ is substantial for some time $\tau$ which causes the qubit to rotate about an axis depending on both $J$ and $\Delta_{SO}$ at a frequency $f = \sqrt{(J(\epsilon))^2+\Delta_{SO}^2}$. Here the qubit is susceptible to charge noise and, as a consequence, begins to dephase. A $\pi$ pulse about the X-axis flips the spin across the Bloch sphere where, upon returning to detuning $\epsilon$, the dephased qubit states refocus for a time $\tau'$. A final $\pi/2$-pulse around the X-axis returns the qubit to the ST basis and a fast adiabatic return pulse projects the states onto the $S(2,0)$ and $T_{0}(1,1)$  basis for measurement. b) A Hahn-echo return for several $\tau$ values along with fits to a Gaussian envelope function. c) Hahn-echo amplitude as a functions of total time evolving under the effect of charge noise ($\tau' +\tau$) The dashed line is a fit to an exponential decay.}
	\label{fig:Supp5}
\end{figure}

\begin{figure}[!h]
	\centering
	\includegraphics[width=0.65\textwidth]{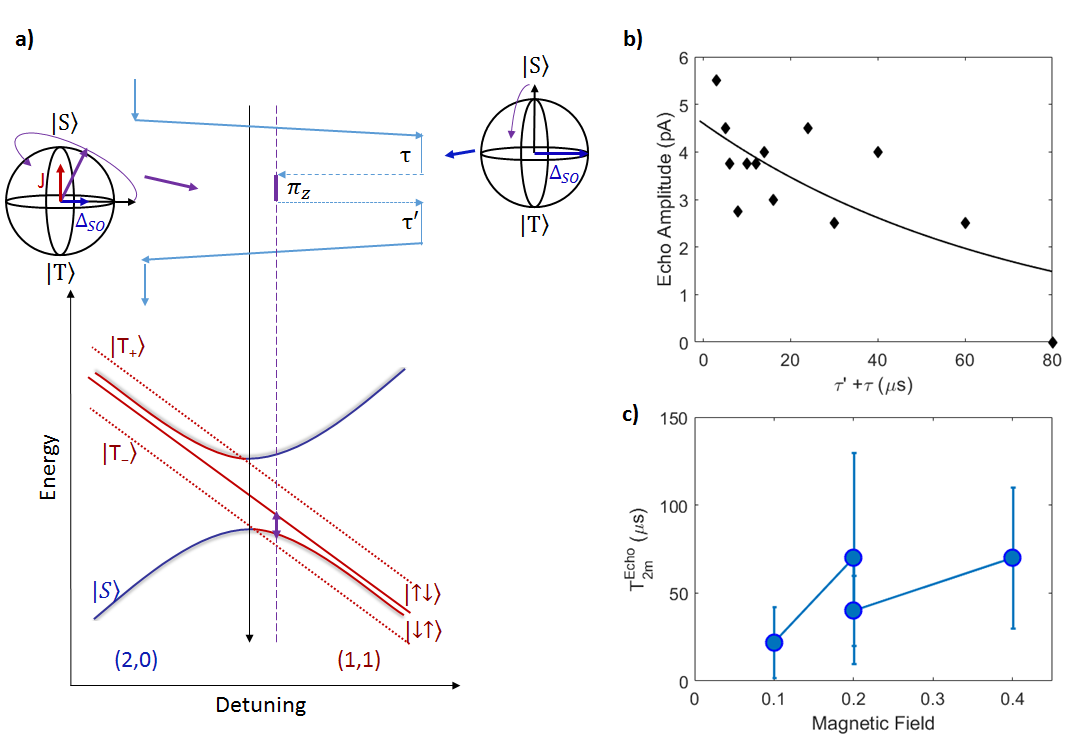}
	\caption{\textbf{Magnetic Noise Hahn-Echo}. a) The qubit is initialized in the S(2,0) ground state and a fast adiabatic pulse transfers the system to the (1,1) charge sector such that it remains in a singlet state where the state is allowed to evolve for some time, $\tau$, about the X-axis under the influence of noise from magnetic fluctuations. A pulse to and from a detuning, $\epsilon$, where $J$ is substantial for a time corresponding to a $\pi$ rotation about the axis depending on both $J$ and $\Delta_{SO}$ flips the spin across the Bloch sphere. The qubit states then evolve again for a time $\tau’$, refocusing the dephased qubit states. A fast adiabatic return pulse projects the states onto the $S(2,0)$ and $(T_{0}(1,1)$  basis for measurement. b) Hahn-echo amplitude as a functions of total time ($\tau + \tau’$) at a magnetic field of 0.2 T along the [100] crystallographic direction.  The line is a fit to an exponential decay. c) The extracted $T_{2m}^{echo}$ for several magnetic field values along the [100]  direction.}
	\label{fig:Supp6}
\end{figure}

The experiments presented in this work indicate that the dephasing of the qubit during rotations is predominantly due to low-frequency, quasi-static noise. Dynamical decoupling techniques may be used to prolong qubit coherence.  This method effectively filters the noise, such that qubit dephasing is most sensitive to noise around the experimental manipulation time. In this work we use a Hahn echo technique \cite{Hahn1950} to decouple from low frequency charge and magnetic noise. To perform a charge noise Hahn echo, a pulse sequence as detailed in Ref. \cite{Dial2013} and depicted in Fig. S\ref{fig:Supp5}(a) is used. For the results presented in the main text, we operated the qubit at a B-field of 0.141 T along the [110] direction giving an X-rotation frequency of $\Delta_{\mathrm{SO}}/h$ = 2.03 MHz. We investigate the effect of a Hahn echo pulse sequence on charge noise decoherence at a detuning where the qubit rotation frequency is 2.24 MHz, corresponding to $J/h$ =  0.99 MHz ($f = \sqrt{J^2+\Delta_{\mathrm{SO}}^2}$).  In Fig. S\ref{fig:Supp5}(b) the measured echo signal is plotted as a function of the difference in evolution times for the first and second $J$-pulse ($\tau'$ - $\tau$) for several total evolution times ($\tau' +\tau$). Here we have subtracted the background charge sensor current, leaving the echoed signal. The echo displays oscillations at 2.24 MHz and an overall Gaussian envelope corresponding to the inhomogeneous dephasing time. $T_{2}^{*}$. By fitting the envelope to the form  $A\exp[-((\tau' - \tau')/T_{2}^{*})^2]$, we can extract the echo amplitude, A, and the dephasing time, $T_{2}^{*}$. We find an average dephasing time $T_{2}^{*}$ = 1.02 $\pm$ 0.06 $\mu$s. In Fig. S\ref{fig:Supp5}(b) we plot the extracted echo amplitude as a function of the total evolution times ($\tau+\tau'$). The data reveals a clear decay in amplitude with a characteristic $1/e$ decoherence time of $T_{2e}^{\mathrm{echo}}$ $\sim$ 8.4 $\mu$s. This is comparable to results observed in GaAs/AlGaAs\cite{Dial2013} and Si/SiGe\cite{Eng2015} ST qubits. 
Similar pulse sequences may be used to decouple the qubit from low-frequency magnetic noise. As shown in Fig. S\ref{fig:Supp6}(a), we use a $\pi$ pulse about the combined $J$ and $\Delta_{\mathrm{SO}}$ axis to create a Hahn-echo. In Fig. S\ref{fig:Supp6}(b) the measured echo signal is plotted as a function of total evolution time under $\Delta_{\mathrm{SO}}$, $\tau' +\tau$, for a B = 0.2 T along the [100] crystallographic direction ($\Delta_{\mathrm{SO}}/h$ $\sim$ 0.5 MHz). An exponential fit reveals a 1/e decay time of $T_{2m}^{\mathrm{echo}}$ $\sim$ 70 $\mu$s. The measured $T_{2m}^{\mathrm{echo}}$  for several magnetic field strengths along the [100] is plotted in Fig. S\ref{fig:Supp6}(c). This value is shorter than other times reported for $T_{2m}^{\mathrm{echo}}$ in silicon\cite{Eng2015}. $T_{2}^{\mathrm{echo}}$ may be bounded by excitation to higher energy states or other $T_{1}$ processes, and further experiments are required to reveal the limiting mechanism. However, this result illustrates our ability to extend coherence times through dynamical decoupling and demonstrates our full two-axis control over the qubit.

\end{document}